\documentclass[fleqn,usenatbib]{mnras}

\usepackage[T1]{fontenc}

\DeclareRobustCommand{\VAN}[3]{#2}
\let\VANthebibliography\thebibliography
\def\thebibliography{\DeclareRobustCommand{\VAN}[3]{##3}\VANthebibliography}

\usepackage{graphicx}	
\usepackage{amsmath}	
\usepackage{amssymb}	
\usepackage{xcolor}
\usepackage{gensymb}
\usepackage{newtxtext,newtxmath}

\newcommand{\Msun}{\mathrm{M_{\odot}}}
\newcommand{\fable}{\textsc{fable}}

\title[AGN feedback in dwarf galaxies]{A little FABLE: exploring AGN feedback in dwarf galaxies with cosmological simulations}

\author[S. Koudmani et al.]{
Sophie Koudmani,$^{1,2}$\thanks{E-mail: skoudmani@ast.cam.ac.uk}
Nicholas A. Henden$^{1}$
and Debora Sijacki$^{1,2}$
\\
$^1$ Institute of Astronomy, University of Cambridge, Madingley Road, Cambridge CB3 0HA, UK \\
$^{2}$ Kavli Institute for Cosmology, University of Cambridge, Madingley Road, Cambridge CB3 0HA, UK
}

\date{Accepted XXX. Received YYY; in original form ZZZ}

\pubyear{2021}

\begin{document}
\label{firstpage}
\pagerange{\pageref{firstpage}--\pageref{lastpage}}
\maketitle

\begin{abstract}
Contrary to the standard lore, there is mounting observational evidence that feedback from active galactic nuclei (AGN) may also play a role at the low-mass end of the galaxy population. We investigate this using the cosmological simulation suite \fable, with a particular focus on the dwarf regime ($M_\mathrm{stellar} < 10^{9.5} \ \Msun$). We find that overmassive black holes (BHs), with respect to the mean scaling relations with their host galaxies, drive hotter and faster outflows and lead to significantly reduced gas mass fractions. They are also more likely to display a kinematically misaligned ionized gas component in our mock MaNGA velocity maps, although we caution that cosmic inflows and mergers contribute to misalignments as well. While in the local Universe the majority of AGN in dwarfs are much dimmer than the stellar component, for $z \geq 2$ there is a significant population that outshines their hosts. These high-redshift overmassive BHs contribute to the quenching of dwarfs, whereas at late cosmic times supernova (SN) feedback is more efficient. While our results are overall in good agreement with X-ray observations of AGN in dwarfs, the lack of high-luminosity X-ray AGN in \fable \ at low redshifts highlights an interesting possibility that SN feedback could be too strong in \fable's dwarfs, curtailing AGN growth and feedback. We predict that future observations may uncover many more AGN in dwarfs with lower luminosities and at higher redshifts.
\end{abstract}

\begin{keywords}
methods: numerical -- galaxies: formation -- galaxies: evolution -- galaxies: dwarf -- galaxies: active
\end{keywords}



\section{Introduction} \label{sec:intro}
In the $\Lambda$CDM model of structure formation, primordial density fluctuations grow and collapse into dark matter (DM) haloes. The gas collapses within these haloes, cools and condenses to form stars. The dark and baryonic components then build up together by accreting and merging with other structures. However, when comparing the DM halo mass function to the galaxy stellar mass function (GSMF), both the low-mass end and the high-mass end of the GSMF are found to be significantly suppressed.

Feedback processes are usually invoked to explain this mismatch and to regulate the overall efficiency of star formation. In the common theoretical model, the low-mass end of the GSMF is suppressed by
reionization \citep[e.g.][]{Efstathiou1992,Okamoto2008,Fitts2017} and SN feedback \citep[e.g.][]{Larson1974,Dekel1986,Mori2002}, whilst AGN feedback regulates the high-mass end \citep[e.g.][]{Binney1995,DiMatteo2005a,Bower2006,Croton2006,Sijacki2007,Cattaneo2009}.

At the high-mass end of the galaxy population, simulations have shown that AGN feedback can bring stellar properties into agreement with observations \citep{Puchwein2013,Dubois2014a,Vogelsberger2014,Khandai2015,Schaye2015,Weinberger2017,Henden2018a}. However, whether reionization and SN feedback are sufficient to regulate star formation at the low-mass end is still unclear. Whilst some find that these two processes can suppress star formation in low-mass haloes \citep[e.g.][]{Governato2007,Sawala2016}, other groups find that additional physical processes (e.g. photoionization, radiation pressure, or stellar winds) are required for SNe to couple to the interstellar medium (ISM) effectively and drive outflows \citep[see e.g.][]{Hopkins2014,Kimm2015,Emerick2018,Smith2018b}. Recently, it has been speculated that AGN may also contribute to regulating star formation at the low-mass end \citep[see e.g.][]{Silk2017}. This idea has been motivated by several observational studies which have identified a population of dwarf galaxies hosting AGN.

Some of the first systematic searches for AGN in dwarf galaxies were carried out by searching SDSS dwarf galaxies for optical AGN signatures \citep[see e.g.][]{Greene2004,Greene2007b,Reines2013,Chilingarian2018a}. These studies use the BPT \citep{Baldwin1981a} diagram to confirm the AGN nature of these sources and broad optical emission lines to estimate the BH mass, where available, measuring BH masses down to $M_\mathrm{BH}\sim 3 \times 10^{4} \ \Msun$. Note that \citet{Cann2019} find that standard optical spectral classification schemes used to identify higher mass BHs do not apply when the BH mass falls below $M_\mathrm{BH} \sim 10^{4} \ \Msun$, so BPT diagnostics may become less useful for exploring AGN activity in the low-mass regime. Moreover, low-mass AGN can easily be outshone by star formation in the optical band. Therefore only bright AGN are recovered resulting in low occupation fractions, e.g. \citet{Reines2013} obtain an AGN occupation fraction of $\sim 0.5$ per cent for their optically-selected sample of AGN in dwarf galaxies.

These two issues underline the importance of multi-wavelength studies. For example, X-ray observations have the advantage that they are less biased towards type 1 AGN (especially at higher redshifts), and that they suffer less from contamination from star formation. Follow-up X-ray observations can then be used to confirm optically-selected AGN \citep{Baldassare2015,Baldassare2017,Chilingarian2018a}. Furthermore, wide-field X-ray surveys have identified large samples of AGN in dwarf galaxies via their X-ray emission \citep[e.g.][]{Schramm2013,Lemons2015,Miller2015,Mezcua2016,Mezcua2018,Pardo2016a,Aird2018a,Birchall2020}. These X-ray surveys also include AGN in dwarf galaxies that would have been missed by the BPT diagnostics due to the optical AGN contribution being less than that from galaxy emission \citep[see e.g.][]{Birchall2020} resulting in higher AGN occupation fractions than for the optically-selected samples. For example, \citet{Pardo2016a} find an overall AGN occupation fraction of $0.6 - 3$ per cent, whilst the occupation for high X-ray luminosities ($\log(L_\mathrm{X} \ [\mathrm{erg\,s^{-1}}]) \geq 41.5$) is estimated to be $\sim 0.2 - 0.4$ per cent \citep{Aird2018a,Mezcua2018,Birchall2020}, more akin to the optically-selected samples which are likely dominated by bright sources.

Furthermore, infrared (IR) surveys have been able to identify `optically-hidden' AGN in dwarf galaxies \citep[e.g.][]{Satyapal2014,Sartori2015a,Marleau2017,Kaviraj2019}, resulting in significantly higher AGN fractions than from optical samples, e.g. \citet{Kaviraj2019} find AGN occupation fractions for dwarf galaxies in the range $\sim 10 - 30$ per cent. Though see \citet{Lupi2020} who caution that resolution effects and source overlapping may contaminate IR-selected AGN samples in the low-mass regime.

Radio observations have also been employed to search for AGN in dwarf galaxies, yielding a sample of AGN with radio jets in the low-mass regime \citep[e.g.][]{Greene2006, Wrobel2006a, Wrobel2008, Nyland2012a,Nyland2017a, Reines2012,Reines2014, Mezcua2018, Mezcua2019, Reines2020}.

Resolved observations are another promising tool for revealing moderate-luminosity AGN in dwarf galaxies, which would be missed using integrated emission line diagnostics \citep[e.g.][]{Dickey2019a,Mezcua2020a}.

Finally, there are several future facilities that are expected to find large samples of BHs in dwarf galaxies. Long-term optical variability searches will be instrumental for identifying AGN missed by other selection techniques, especially in the era of LSST \citep[e.g.][]{Baldassare2020}. The gravitational wave observatory LISA will be aimed at detecting gravitational waves from the coalescence of BHs with 
masses $M_\mathrm{BH} \sim 10^{4} - 10^{7} \Msun$ and is therefore optimally placed to study the BH population in the low-mass regime. Recent simulation studies have highlighted the multimessenger signatures of BHs in dwarfs \citep{Bellovary2019,Volonteri2020}. JWST will also provide important insights into AGN in dwarf galaxies and alleviate the above-mentioned resolution issues of current IR studies \citep[e.g.][]{Cann2018a}. The combination of JWST with future X-ray missions (\textit{Athena}, \textit{Lynx}) will be a powerful tool for mapping the AGN population in dwarf galaxies at higher redshifts \citep[e.g.][]{Civano2019,Haiman2019} and lower luminosities \citep[e.g.][]{Schirra2020}. The ngVLA will map the weakly-accreting AGN in dwarf galaxies and provide insights into the jet-driven feedback mechanisms in the low-mass regime as well as BH seeding mechanisms \citep[e.g.][]{Nyland2018,Plotkin2018a}.

Beyond constraining the AGN demographics at the low-mass end, recent observations have started studying whether AGN in dwarf galaxies could have a significant impact on their hosts - with a particular focus on quenching and outflows. \citet{Penny2018a} study quiescent dwarf galaxies in the MaNGA survey and find that six out of 69 quiescent MaNGA dwarfs have optical AGN signatures. Interestingly, five out of six of these dwarfs have an ionized gas component that is kinematically offset from the stellar component which could be a sign of AGN-driven outflows. \citet{Manzano-King2019} find direct evidence of spatially extended high-velocity ionized gas outflows in nine dwarf galaxies hosting AGN, for six of these the emission-line ratios of the outflow itself are consistent with AGN ionization, and \citet{Liu2020} present an integral field spectroscopy follow-up study of a subsample of these dwarf galaxies, investigating the gas outflows in more detail. \citet{Dickey2019a} study quiescent dwarfs in isolated environments and find that 16 out of 20 quiescent isolated dwarfs have AGN-like line ratios. 

These recent observations of AGN activity in dwarf galaxies have motivated theorists to investigate this largely unexplored regime - with mixed results. On the one hand, analytical models of AGN feedback in dwarf galaxies look promising, suggesting that AGN could provide an alternative and more successful source of negative feedback than SNe \citep{Silk2017,Dashyan2018}. On the other hand, in numerical simulations, strong SN feedback can hinder the growth of BHs in low-mass galaxies rendering AGN feedback ineffective \citep[see e.g.][]{Dubois2015,Angles-Alcazar2017,Habouzit2017,Trebitsch2018}. However, if BHs in simulations are able to grow efficiently, they can have a significant impact on their hosts \citep{Barai2018b,Sharma2020}. We have systematically tested the effect of AGN activity in a series of isolated simulations of dwarf galaxies \citep{Koudmani2019}. We found that the outflows reach significantly higher velocities and temperatures with added AGN feedback. However, AGN activity did not have a significant effect on the global star formation rate (SFR) in our isolated set-up.

The considerations above suggest an interesting possibility that AGN feedback may play an indirect role in regulating star formation at the low-mass end. For example, the AGN-boosted outflows might be able to hinder cosmic gas inflows. Cosmological simulations are needed to investigate these possibilities.

In this paper, we use the \fable \ (\textit{Feedback Acting on Baryons in Large-scale Environments}) simulation suite \citep{Henden2018a,Henden2019a,Henden2019} to explore the role of AGN feedback in the low-mass regime within a realistic cosmological environment. Keeping in mind that the \fable \ parameters have been tuned so that SN feedback is strong to regulate the low-mass end of the GSMF, we investigate whether low-mass galaxies with efficient BH growth have significantly different physical properties than the ones without AGN activity - with a particular focus on outflow properties and SFRs. Though we note that any effect we find from AGN feedback in the low-mass regime is likely to be a lower limit. Furthermore, we aim to compare the \fable \ galaxies to observations of dwarf galaxies with AGN and to make predictions for future observations.

The outline of this paper is as follows. We describe the characteristics of the \fable \ simulation suite and the sub-grid models in Sections~\ref{subsec:sims}~and~\ref{subsec:feedback}, respectively. The methods for identifying haloes and galaxies in \fable \ and the definition of the low-mass galaxy sample are given in Section~\ref{subsec:galid} and we describe how we calculate the outflow properties of these galaxies in Section~\ref{subsec:OutflowMethods}. Our methods for creating mock MaNGA observations and X-ray luminosities are outlined in Sections~\ref{subsec:MaNGAMethods}~and~\ref{subsec:Xraymethods}. In Section~\ref{sec:results}, we present our results on the AGN population (Section~\ref{subsec:population}), the scaling relations (Section~\ref{subsec:scalings}), the properties of low-mass galaxies with overmassive BHs (Section~\ref{subsec:offsetBHs}), mock MaNGA observations (Section~\ref{subsec:MockMaNGAObs}), and mock X-ray observations (Section~\ref{subsec:xrayprops}). We discuss our results in Section~\ref{sec:discussion} and summarise our main findings in Section~\ref{sec:conclusions}.

\section{Methodology} \label{sec:methods}
\subsection{Basic simulation properties} \label{subsec:sims}
We use the cosmological simulation suite \fable \  for our analysis of AGN feedback in low-mass galaxies. The \fable \  simulation suite is extensively described in \citet{Henden2018a}, and only a brief summary is given below.

The \fable \  simulations were carried out with the massively parallel \textsc{arepo} code \citep{Springel2010}, where fluid dynamics is discretized on a moving mesh using a quasi-Lagrangian finite volume technique. The unstructured mesh is based on the Voronoi tessellation of a set of discrete points that cover the whole computational domain. These mesh-generating points are allowed to move with the local flow velocity, with minor corrections to avoid excessive distortion of the gas cells. Gravitational interactions are modelled using the TreePM approach, with stars and DM modelled as collisionless particles.

The \fable \  suite consists of a large-volume cosmological box simulation as well as a series of zoom-in simulations of individual groups and clusters. As we are interested in low-mass galaxies in representative regions of the Universe, we focus only on the results from the cosmological box for this study.

The comoving 40~$h^{-1}$~Mpc box was evolved using initial conditions for a uniformly-sampled cosmological volume based on a Planck cosmology \citep[see][]{PlanckCollaboration2016}. The volume contains $512^3$ DM particles with masses $m_\mathrm{DM} = 3.4 \times 10^7 ~ h^{-1}\, \Msun$ and initially $512^3$ gas resolution elements with target mass $\overline{m}_\mathrm{gas} = 6.4 \times 10^{6} ~ h^{-1}\, \Msun$. The gravitational softening length is set to $2.393 ~ h^{-1} \, \mathrm{kpc}$ in physical coordinates below $z=5$ and fixed in comoving coordinates at higher redshift.

The \fable \  sub-grid models are largely based on the Illustris galaxy formation model \citep{Vogelsberger2013a,Vogelsberger2014,Genel2014,Torrey2013,Sijacki2015a}. Whilst the models for star formation \citep{Springel2003a}, radiative cooling \citep{Katz1996,Wiersma2009a}, and chemical enrichment \citep{Wiersma2009} are unchanged from Illustris, the stellar feedback \citep{Vogelsberger2013a} and AGN feedback \citep{DiMatteo2005a,Springel2005,Sijacki2007,Sijacki2015a} models have been updated (see Section~\ref{subsec:feedback}). Both the stellar and the AGN feedback models were calibrated to reproduce observations of the local GSMF and the gas mass fractions of observed massive haloes. 

\subsection{Feedback models} \label{subsec:feedback}

\subsubsection{Stellar feedback}
Stellar feedback is included as a sub-grid model for galactic winds and outflows \citep[see][for details]{Vogelsberger2013a}. In the Illustris simulation, these winds are purely kinetic. In \fable, this is modified so that one third of the wind energy is imparted as thermal energy, which slows down the cooling of the gas thereby increasing the overall effectiveness of the stellar feedback \citep[see][]{Marinacci2014, Henden2018a}.

\subsubsection{BH seeding and growth} \label{subsubsec:SeedingAndGrowth}
BHs are modelled as collisionless sink particles. BH seeds of mass $M_\mathrm{BH,seed}= 10^{5} \ h^{-1}\, \Msun$ are placed into every DM halo above $5\times 10^{10} \ h^{-1} \, \Msun$. Subsequently, these seeds can grow via mergers with other BHs and via gas accretion \citep[see][for details]{Vogelsberger2013a}.
In brief, BH particles accrete at the Eddington-limited Bondi-Hoyle-Lyttleton-like rate boosted by a factor of $\alpha=100$ \citep{Hoyle1939,Bondi1944,Springel2005}. The radiative efficiency is set to be a constant $\epsilon_\mathrm{r}=0.1$. A fraction of $(1-\epsilon_\mathrm{r})$ of the accreted mass is added to the BH mass, whilst the rest is available as feedback energy.

Following Illustris, \fable \ uses a repositioning scheme to ensure that the BH particle stays at the gravitational minimum of its host halo. This prevents spurious BH movement due to two body scattering effects with massive DM or star particles \citep[see][]{Sijacki2015a}. Though we note that this also precludes us from investigating physical off-nuclear BH motion as has been found in some observations \citep[e.g.][]{Mezcua2020a,Reines2020} and simulations \citep[e.g.][]{Bellovary2019} of dwarf galaxies. However, there are currently only few firm observational indications of offset massive BHs \citep[e.g.][]{Komossa2012,Shen2019}. Also see \citet{Tremmel2018} for a discussion of off-nuclear BHs in more massive objects.

\subsubsection{AGN feedback} \label{subsubsec:AGNfeed}
In analogy to Illustris, the AGN feedback operates in either one of two states depending on the Eddington fraction $f_\mathrm{Edd}$: the high accretion rate quasar mode \citep{DiMatteo2005a,Springel2005} or the low accretion rate radio mode \citep{Sijacki2007}. The BH particles switch to the quasar mode once they exceed a critical Eddington fraction $f_\mathrm{Edd,QM}$. In the \fable \  model, this threshold is set to $f_\mathrm{Edd,QM}=0.01$.

This AGN feedback model has been modified in \fable \  by introducing a duty cycle to the quasar mode. The feedback energy is accumulated over $\delta t = 25 \ \mathrm{Myr}$, before being released in a single event, to reduce overcooling \citep[see][]{Booth2009,Henden2018a}. 

The remainder of the feedback energy that is not coupled to the surrounding gas by the quasar or radio mode goes into radiative electromagnetic feedback, which is approximated as an additional radiation field around the BH superposed with the redshift-dependent ultraviolet background \citep{Vogelsberger2013a}.

\subsection{Galaxy identification} \label{subsec:galid}
Haloes and subhaloes are identified via friends-of-friends (FoF) and \textsc{subfind} algorithms \citep{Davis1985, Springel2001,Dolag2009}. For the FoF search, we choose a linking length of $0.2$ multiplied by the mean inter-particle separation. We then use \textsc{subfind} to identify gravitationally self-bound subhaloes within each FoF halo. The `central subhalo' is the subhalo at the minimum gravitational potential of the FoF halo. All other subhaloes in the FoF halo are categorised as satellite subhaloes.

We consider each subhalo with at least 100 star particles to be a resolved galaxy, yielding a minimum stellar mass of approximately $10^{9}\ \Msun$. For this investigation, we only consider central subhaloes, as our focus is on intrinsic feedback processes rather than environmental effects.

At each redshift, our low-mass galaxy sample consists of all central subhaloes that host a BH and have a stellar mass in the range $9.0 \leq \log(M_\mathrm{stellar} \ [\Msun]) < 10.5$ at the given redshift. Stellar mass here is defined as the mass of all star particles within twice the stellar half mass radius, $r_\mathrm{eff}$. The minimum mass is set by the above resolution requirements. In addition, we restrict our sample to galaxies with masses below the knee of the GSMF ($\mathcal{M^{*}} = 10^{10.5} \ \Msun$), where AGN feedback had previously been assumed to be ineffective.

To investigate the impact of AGN feedback as a function of stellar mass, we divide up the low-mass galaxy sample into three mass bins of equal widths in log space: dwarf galaxies ($9.0 \leq \log(M_\mathrm{stellar} \ [\Msun]) < 9.5$), massive dwarf galaxies ($9.5 \leq \log(M_\mathrm{stellar} \ [\Msun]) < 10.0$), and $\mathcal{M^{*}}$ galaxies ($10.0 \leq \log(M_\mathrm{stellar} \ [\Msun]) < 10.5$).

\subsection{Outflow properties} \label{subsec:OutflowMethods}
We would like to assess whether AGN feedback can have an impact on the overall outflow properties of low-mass galaxies. To this end, we extract the outflow properties from the simulation data for the whole gas (i.e. not separating out the different gas phases). Though note that \fable \ does not include cooling below $10^{4} \ \mathrm{K}$, so these outflow properties mainly correspond to the warm and to the hot gas phase. 

We obtain the outflow properties for each galaxy as follows: Firstly, we centre the gas coordinates on the potential minimum of the galaxy. Then we subtract the peculiar velocity (here defined as the mass-weighted mean velocity of all gas cells within twice the virial radius\footnote{Throughout this paper, the virial quantities of a system are defined by a mean density of $200\times$ the critical density of the Universe.}, $2R_\mathrm{vir}$) from the gas velocities. We then calculate the outflow properties at two different target radii $r_\mathrm{t} = 0.2R_\mathrm{vir}, R_\mathrm{vir}$ within shells of widths $\Delta r = 10, 20 \ \mathrm{kpc}$, respectively\footnote{Note that in both cases we tested various shell widths and found that the results are well converged and do not depend on the exact choice of $\mathrm{\Delta}r$, as long as the width is large enough to have a sufficient number of gas cells within the shell.}. We select all gas cells within the $r_\mathrm{t} \pm \frac{\Delta r}{2}$ shell and calculate the radial velocities $v_{r}$ for all of these cells. We then mark gas cells with $v_{r}<0$ as inflowing and gas cells with $v_{r}>0$ as outflowing. Note that we only calculate the outflow properties for galaxies that have at least ten gas cells (at least five inflowing and at least five outflowing) within the shell considered.

We then use the inflowing and outflowing gas cells to calculate the mass flow rates through the $r_\mathrm{t} \pm \frac{\Delta r}{2}$ shell as follows:
\begin{equation}
    \dot{M}_\mathrm{out} = \sum_{\substack{|r-r_\mathrm{t}|< \frac{\Delta r}{2} \\ v_{r} >0}} \frac{m_\mathrm{cell} v_{r}}{\Delta r},
\end{equation}

\begin{equation}
    \dot{M}_\mathrm{in} = \sum_{\substack{|r-r_\mathrm{t}|< \frac{\Delta r}{2} \\ v_{r} < 0}} \frac{m_\mathrm{cell} v_{r}}{\Delta r},
\end{equation}

\begin{equation}
    \dot{M}_\mathrm{tot} = \dot{M}_\mathrm{out} + \dot{M}_\mathrm{in}.
\end{equation}

Here $m_\mathrm{cell}$ is the gas cell mass, $\dot{M}_\mathrm{out}$ is the outflow rate, $\dot{M}_\mathrm{in}$ is the inflow rate, and $\dot{M}_\mathrm{tot}$ is the total mass flow rate.

We also calculate the outflow velocities and temperatures. We define the outflow velocity, $v_\mathrm{out}$, as the mass-weighted mean velocity of all gas cells within the shell centred at $r_\mathrm{t}$ with $v_{r} > 0$. Similarly, we define the outflow temperature, $T_\mathrm{out}$, as the mass-weighted mean temperature of all cells within the shell centred at $r_\mathrm{t}$ with $v_{r} > 0$. 

Using the above quantities, we then also calculate the momentum outflow rate, $\dot{P}_\mathrm{out}$, and the kinetic energy outflow rate, $\dot{E}_\mathrm{kin, out}$, as: 
\begin{equation}
    \dot{P}_\mathrm{out} = \sum_{\substack{|r-r_\mathrm{t}|< \frac{\Delta r}{2} \\ v_{r} >0}} \frac{m_\mathrm{cell} v_{r}}{\Delta r} v_{r},
\end{equation}
\begin{equation}
    \dot{E}_\mathrm{kin, out} = \frac{1}{2} \sum_{\substack{|r-r_\mathrm{t}|< \frac{\Delta r}{2} \\ v_{r} >0}} \frac{m_\mathrm{cell} v_{r}}{\Delta r} v_{r}^{2}.
\end{equation}
Finally, we also calculate the mass loading factor, $\beta$, as 
\begin{equation}
\beta = \frac{\dot{M}_\mathrm{out}}{\mathrm{SFR}(r \leq r_\mathrm{t})},
\end{equation}

where $\mathrm{SFR}(r \leq r_\mathrm{t})$ is the summed up SFR of all gas cells within $r_\mathrm{t}$.

\subsection{Mock MaNGA maps} \label{subsec:MaNGAMethods}
The MaNGA survey found kinematically misaligned gas in quenched dwarf galaxies with AGN signatures hinting at AGN-driven outflows \citep[see][]{Penny2018a}. In \citet{Koudmani2019}, we created mock MaNGA maps of isolated simulations of dwarf galaxies to explore these kinematic offsets. Here we adapt this procedure to cosmological simulations, as described below.

\subsubsection{Sample selection}
To accurately determine the kinematic position angle (PA), the galaxies need to have clear rotational features. To automatically select galaxies that are rotationally supported, we only keep galaxies with $V_\mathrm{max}/ \sigma > 1.1$, where $V_\mathrm{max}$ is the maximum value of the respective subhalo's spherically averaged rotation curve and $\sigma$ is the 3D velocity dispersion of the same subhalo. Visual inspection of our sample galaxies confirms that this is an efficient criterion for ensuring sufficient rotational support.
 
All star particles or ionized gas cells that belong to the galaxy's FoF group are included along the projection direction for the mock line-of-sight (l.o.s.) velocity maps. We require that there should be at least 50 star particles and 50 ionized gas cells within the $1.5r_\mathrm{eff}$ 2D aperture (corresponding to the spatial coverage of the primary MaNGA sample). Here $r_\mathrm{eff}$ is the stellar half-mass radius. This ensures that the kinematics are well resolved by the simulation. Due to the weightings (see Section~\ref{subsubsec:losmaps}) the projected l.o.s. velocities are mostly sensitive to the kinematics of the central galaxy and are measured up to the scale of the virial radius, and therefore these measurements are only minimally affected by the (significantly smaller) gravitational softening.
 
\subsubsection{Line-of-sight velocity maps} \label{subsubsec:losmaps}

We match the resolution of our mock MaNGA maps to the pixel size of the MaNGA final reduced data cubes (0.5 arcsec). We analyse three different redshifts ($z=$ 0, 0.2, 0.4). Note that for the local $z=0$ sample, we calculate the spatial resolution assuming that these galaxies are observed at the mean redshift of the primary MaNGA sample, $\bar{z} = 0.03$. We do not consider higher redshifts than $z=0.4$, as the resolution of the MaNGA survey would prohibit us from confidently identifying kinematic PAs.

We create separate l.o.s velocity maps for the star particles and the ionized gas cells. We keep all gas cells with neutral hydrogen abundance, fraction of the hydrogen cell mass (or density) in neutral hydrogen, $n_\mathrm{H, neutral} < 10^{-2}$ to isolate the ionized gas\footnote{We plotted the $n_\mathrm{H, neutral} - v_\mathrm{l.o.s.}$ distribution for a sample of galaxies across different mass bins and redshifts, and found that there was a clear break below that value.}.

For each subhalo, we centre the velocity map on the minimum gravitational potential and subtract the systemic velocity, here taken as the mass-weighted mean velocity within $1.5r_\mathrm{eff}$. 

To separate out the effect of the inclination angle on kinematic misalignments, we rotate all galaxies into the same orientation for our analysis. To this end, we calculate the stellar total angular momentum $\mathbf{L}_\mathrm{*}$ within $1.5r_\mathrm{eff}$ and rotate each galaxy (both stars and gas) so that $\mathbf{L}_\mathrm{*}$ is aligned with the z-axis. We then project the stellar / gas kinematics onto the y-z plane so that all galaxies are viewed edge-on. As an optional additional step, the disc can then be inclined by rotating about the y-axis to test the effect of the inclination angle.

We project the ionized gas / stellar velocity distribution onto a grid corresponding to the spatial extent and resolution of the MaNGA data. We define the stellar smoothing length as the radius enclosing the 64 nearest neighbours and the gas smoothing length as $2.5$ times the cell radius. We set a maximum smoothing length as $50$ times the pixel size, though in practice that limit is rarely reached.

We weight the velocity contributions to each pixel by the cubic spline smoothing kernel. In addition, we impose a luminosity-like weighting: the stars by their mass\footnote{We repeated these measurements for a subset of our galaxy sample weighting the star particle contributions by the stellar bolometric luminosity instead of by the stellar mass. We found that the results regarding the kinematic PA are not affected but that the velocity maps look somewhat different because the rotationally supported stars enter with different weights.} and the gas cells by the square of their density. Note that when creating these smoothed maps we consider all resolution elements within a $2.5r_\mathrm{eff}$ 2D aperture as some might contribute to pixels within the $1.5r_\mathrm{eff}$ 2D aperture even though their centres lie outside that aperture. 

Finally, we convolve the resulting velocity map with a Gaussian filter of $\mathrm{FWHM} = 2.5$ arcsec to model the effect of the MaNGA PSF.

\subsubsection{Kinematic position angles}

The kinematic PA is related to the projected angular momentum vector and traces the position of the maximum change in velocity on the map \citep[see e.g.][]{Krajnovic2006}. Following \citet{Penny2018a}, we determine the kinematic PAs of both ionized gas and stars using the \texttt{fit\_kinematic\_pa} routine \citep[see][for a detailed description]{Krajnovic2006}. Note that this routine only returns PAs in the range $0 \degree \leq \mathrm{PA} < 180 \degree$, which does not take the sense of the rotation into account. We identify the redshifted part of the velocity map to convert the PAs to the $0 \degree \leq \mathrm{PA} < 360 \degree$ range.

The kinematic offset between gas and stars is then defined as $\Delta \mathrm{PA} = |\mathrm{PA}_\mathrm{stars} - \mathrm{PA}_\mathrm{gas}|$. Gas and stars are considered to be in dynamical equilibrium if $\Delta \mathrm{PA} =$ 0\degree or 180\degree. As in the observational study by \citet{Penny2018a}, we then define the gas as kinematically misaligned if 30\degree $\leq \Delta \mathrm{PA} <$ 150\degree.

\subsection{Mock X-ray luminosities}
\label{subsec:Xraymethods}

X-ray studies by \textit{Chandra} and \textit{XMM-Newton} have resulted in large samples of AGN candidates in low-mass galaxies. As X-ray surveys push to higher redshifts, it has become possible to start studying the redshift evolution of AGN activity in these galaxies. To compare the AGN population in \fable \  to these observational findings, we calculate the X-ray properties of the \fable \  galaxies.

We define the bolometric BH luminosity as $L_\mathrm{BH} = \epsilon_\mathrm{r} \dot{M}_\mathrm{BH} \mathrm{c}^{2}$, where $\epsilon_{r}$ is the radiative efficiency, $\dot{M}_\mathrm{BH}$ is the BH accretion rate and $\mathrm{c}$ is the speed of light (see Section~\ref{subsubsec:AGNfeed}). To obtain the BH luminosities in the soft (0.5 - 2 keV) and hard (2 - 10 keV) X-ray band, we use the bolometric corrections from \citet{Shen2020}. To assess the detectability of these low-luminosity AGN, we also estimate the X-ray luminosities of X-ray binaries (XRBs) and the hot gas in the galaxy.

To obtain an estimate for the contribution from XRBs, we use the redshift-dependent relation from \citet{Lehmer2016} which relates the total X-ray luminosity $L_\mathrm{X,XRB}$ to the SFR and stellar mass ($M_\mathrm{stellar}$):
\begin{equation}
    \left( \frac{L_\mathrm{X,XRB}}{\mathrm{erg\,s^{-1}}} \right) = \alpha_{0} (1+z)^{\gamma} \left(\frac{M_\mathrm{stellar}}{\Msun}\right) + \beta_{0} (1+z)^{\delta} \left(\frac{\mathrm{SFR}}{\mathrm{\Msun\, yr^{-1}}} \right).
\end{equation}
The parameter values in the soft X-ray band are $\log(\alpha_{0})=29.04$, $\log(\beta_{0})=39.38$, $\gamma = 3.78$, and $\delta=0.99$. In the hard X-ray band, the parameter values are given by $\log(\alpha_{0})=29.37$, $\log(\beta_{0})=39.28$, $\gamma = 2.03$, and $\delta=1.31$. To estimate the \fable \ luminosities, we evaluate both the SFR and $M_\mathrm{stellar}$ within $2r_\mathrm{eff}$.

We calculate the hot gas contribution using the relationship between the X-ray luminosity of the diffuse ISM and the SFR from \citet{Mineo2012}:
\begin{equation}
    \left( \frac{L_\mathrm{X,gas}}{\mathrm{erg\,s^{-1}}} \right) = 8.3 \times 10^{38} \times \left( \frac{\mathrm{SFR}}{\mathrm{\Msun\, yr^{-1}}} \right).
\end{equation}
Again, we calculate the SFR within $2r_\mathrm{eff}$ here. Note that the above relation only applies to soft X-ray luminosities. To obtain the contribution in the hard X-ray band, we assume a photon index $\Gamma = 3$, which is a good representation of a thermal model with temperature 0.7 - 1 keV \citep[see][]{Mezcua2018}.

\section{Results} \label{sec:results}
\subsection{The AGN population} \label{subsec:population}
\begin{figure*}
    \centering
    \includegraphics[width=\textwidth]{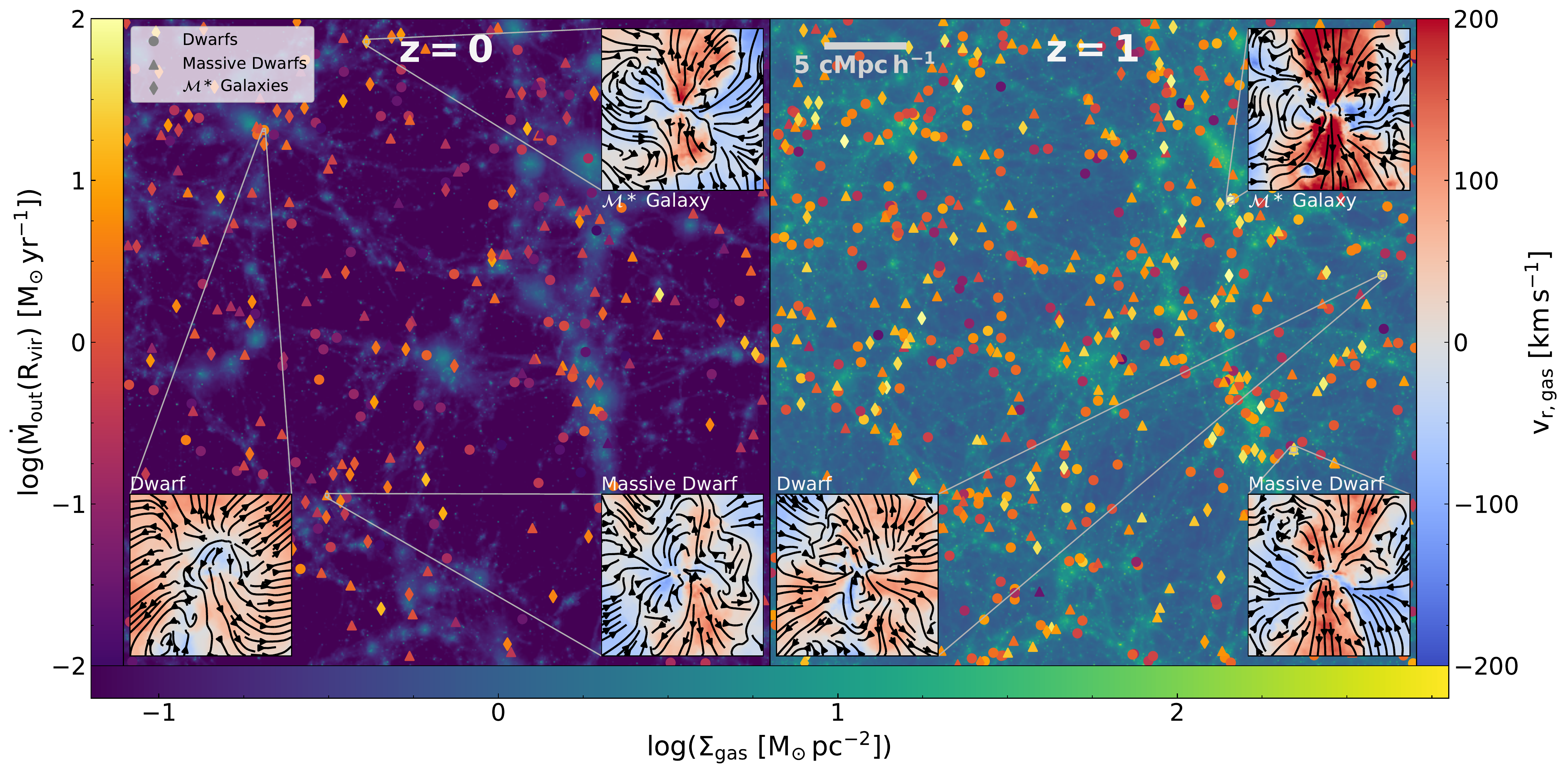}
    \caption{Surface density projections of the total gas for the whole \fable \ simulation box at redshifts $z=0$ (left panel) and $z=1$ (right panel). The markers show the locations of low-mass galaxies that host an actively accreting BH ($f_\mathrm{Edd}>10^{-3}$). The marker style indicates the respective stellar mass bin with the sample divided up into dwarf galaxies ($9.0 \leq \log(M_\mathrm{stellar} \ [\Msun]) < 9.5$), massive dwarf galaxies ($9.5 \leq \log(M_\mathrm{stellar}\ [\Msun]) < 10.0$), and $\mathcal{M^{*}}$ galaxies ($10.0 \leq \log(M_\mathrm{stellar}\ [\Msun]) < 10.5$). The marker colour shows the mass outflow rate of the total gas at the virial radius, $R_\mathrm{vir}$. For each of the three mass bins at both redshifts, we select an example galaxy with a high outflow rate and show the projected outflow kinematics of the total gas in the zoomed-in insets (projection dimensions $2R_\mathrm{vir} \times 2R_\mathrm{vir} \times 2R_\mathrm{vir}$). The inset colour-coding indicates the radial velocity and the velocity streamlines are shown in black. Note that these insets have been rotated so that the galaxies are viewed edge-on.}
    \label{fig:box_projection}
\end{figure*}

We start our analysis by inspecting the large-scale distribution of AGN in low-mass galaxies in our simulation. Figure~\ref{fig:box_projection} shows projections of the total gas surface density for the whole \fable \  simulation box at $z=0$ (left panel) and $z=1$ (right panel). The markers show the locations of active low-mass galaxies (with Eddington fractions $f_\mathrm{Edd}>10^{-3}$). The galaxies' stellar masses are indicated by the marker style with the sample split into dwarfs with $9.0 \leq \log(M_\mathrm{stellar} \ [\Msun]) < 9.5$, massive dwarfs with $9.5 \leq \log(M_\mathrm{stellar} \ [\Msun]) < 10.0$, and $\mathcal{M^{*}}$ galaxies with $10.0 \leq \log(M_\mathrm{stellar}\ [\Msun]) < 10.5$. Focusing on the spatial distribution, there are low-mass galaxies with AGN in all types of cosmic environments: voids, filaments, and knots. This is in line with the observational findings from SDSS galaxies, which suggest that environment is not an important factor in triggering AGN activity in dwarf galaxies \citep{Kristensen2020}. There are significantly more active low-mass galaxies at redshift $z=1$ than at $z=0$.

The marker colour shows the galactic outflow rates at the virial radius, $R_\mathrm{vir}$ for the total gas (see Section~\ref{subsec:OutflowMethods}). Across all mass bins (and at both redshifts), we find active galaxies with large-scale outflows at the virial radius ($\dot{M}_\mathrm{out} \gtrsim 10\ \mathrm{M_{\odot}\, yr^{-1}}$). There are higher outflow rates and velocities at redshift $z=1$ compared to $z=0$. Unsurprisingly, more massive galaxies tend to be associated with stronger outflows, given the larger SFRs and BH masses in those systems. Though the more massive galaxies also have larger escape velocities than the low-mass objects.

For each of the three mass bins at both redshifts, we select a representative example galaxy with a high outflow rate and show $2R_\mathrm{vir} \times 2R_\mathrm{vir} \times 2R_\mathrm{vir}$ slices of the outflows in the zoomed-in insets. The inset colour-coding shows projections of the radial velocity of the total gas with the streamlines overlaid on the projection. Note that the galaxies shown in these projections have been rotated so that they are viewed edge-on. The more massive galaxies have collimated outflows, whilst the outflows for the example dwarf galaxies are more isotropic. For all of the example galaxies shown here, the outflows are able to propagate to the virial radius.

The outflow velocities of the total gas at $R_\mathrm{vir}$, as defined in Section~\ref{subsec:OutflowMethods}, range from  $\sim 50$ to $150 \ \mathrm{km\,s^{-1}}$. In most cases the outflows decelerate between $0.2R_\mathrm{vir}$ and $R_\mathrm{vir}$. To estimate the escape velocity for these haloes we use the virial velocity, $V_\mathrm{vir}$, for simplicity, which ranges from $\sim 80$ to $210 \ \mathrm{km\,s^{-1}}$. For most of these example galaxies, the mean outflow velocity at $R_\mathrm{vir}$ does not exceed the virial velocity, indicating galactic fountain nature of the outflows. The exception to this is the dwarf example galaxy at $z=0$. For this galaxy, the outflows accelerate from $42 \ \mathrm{km\,s^{-1}}$ at $0.2R_\mathrm{vir}$ to $94 \ \mathrm{km\,s^{-1}}$ at $R_\mathrm{vir}$, exceeding $V_\mathrm{vir}=81 \ \mathrm{km\,s^{-1}}$. Note that this galaxy has an overmassive BH relative to its stellar mass  with $M_\mathrm{BH}=4\times10^{5}\ \Msun$ and  $M_\mathrm{stellar}=10^{9}\ \Msun$ (see \fable \ scaling relations in Figure \ref{fig:ScalingRelations}) and is in a crowded region with strong cosmic inflows.

Even though the outflow velocities of the total gas do not exceed $V_\mathrm{vir}$ in the other cases shown, the fast phase of the outflow is able to reach velocities higher than $V_\mathrm{vir}$ in all cases. The maximum velocities of the outflowing gas range from $\sim150$ to $500 \  \mathrm{km\,s^{-1}}$. We also checked the temperature distributions of these outflows and find that the outflows at $R_\mathrm{vir}$ are multiphase with temperatures ranging from $8 \times 10^{3} \ \mathrm{K}$ to $3 \times 10^{6} \ \mathrm{K}$. The fast phase of the outflow corresponds to the hot phase, with temperatures between $\sim 10^{5}$~K and $10^{6}$~K.

\begin{figure*}
    \centering
    \includegraphics[width=\textwidth]{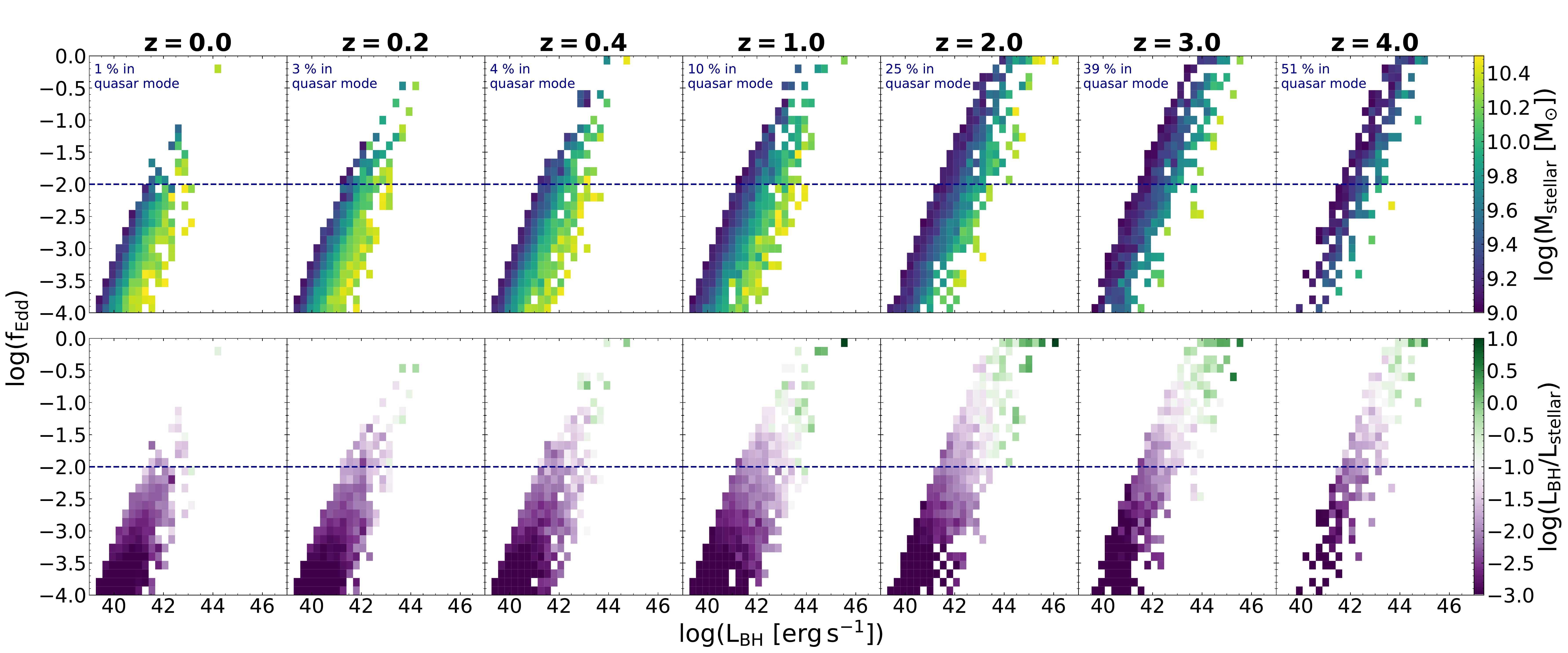}
    \caption{The redshift evolution of the population of BHs hosted by low-mass galaxies ($9.0 \leq \log(M_\mathrm{stellar} \ [\Msun]) < 10.5$). At each redshift, the distribution of BHs in the $L_\mathrm{BH}-f_\mathrm{Edd}$ space is shown. The colour coding indicates the average $M_\mathrm{stellar}$ (top panel) and the average $L_\mathrm{BH}/L_\mathrm{stellar}$ ratio (bottom panel). The Eddington fraction where the BHs transition to quasar mode feedback ($f_\mathrm{Edd,QM}=10^{-2}$) is shown by the blue dashed line. The percentage of BHs that are in the quasar mode (out of the whole low-mass galaxy population) is given in the upper left corner for each redshift. There are very few bright AGN in low-mass galaxies at low redshifts, rendering these AGN hard to detect. The high-redshift population is much more promising  with $\sim 25 - 50$ per cent in the quasar mode. The bottom panel illustrates how in the majority of cases, the stars in the galaxy outshine the AGN, demonstrating the importance of multi-wavelength studies. Again this problem is alleviated for high Eddington fraction accretors at higher redshifts.}
    \label{fig:PopulationStatistics}
\end{figure*}

Having established that a significant number of low-mass galaxies with AGN in \fable \ have appreciable outflows, we move on to investigate the radiative properties of this population. Figure~\ref{fig:PopulationStatistics} shows the distribution of the low-mass AGN population in $L_\mathrm{BH} - f_\mathrm{Edd}$ space, where $L_\mathrm{BH}$ is the bolometric BH luminosity. We show this distribution at a range of redshifts between $z=0$ and $z=4$, focusing on the high Eddington fraction end of this distribution ($f_\mathrm{Edd} \geq 10^{-4}$). The threshold Eddington fraction for transition to the quasar mode ($f_\mathrm{Edd,QM}=10^{-2}$) is shown as a blue dashed line. Furthermore, the fraction of BHs in the quasar mode (out of all BHs hosted by low-mass galaxies) is given in the upper left corner for each redshift.

The colour-coding in the upper panel indicates the average stellar mass in each pixel. Dwarf galaxies (with $\log(M_\mathrm{stellar}\ [\Msun]) < 9.5$) dominate the galaxy sample at high redshifts and the high-z distribution is quite narrow. Moving to lower redshifts, the distribution broadens as we get a wider range of stellar masses.

There is a clear stellar mass gradient where at a fixed BH luminosity, BHs with higher Eddington fractions tend to be hosted by less massive galaxies. There are very few highly-accreting BHs at low redshifts ($z < 1$) rendering these AGN difficult to detect. Though note that \fable \ may underproduce high-luminosity sources at low redshifts (see Section~\ref{subsec:xrayprops}).

At $z=0$, only one per cent of BHs in low-mass galaxies are in the quasar mode. This increases to 10 per cent at $z=1$, and goes up to 51 per cent at $z=4$. Future facilities (such as \textit{Athena}, \textit{Lynx}, JWST or the \textit{Roman Space Telescope}) will be able to probe this early phase of significant activity, as predicted by our simulations.  Note that we have compared the evolution of the number density of AGN for different luminosity bins in \fable \ with \citet{Hopkins2007a}. We found that this is in good agreement and reproduces the cosmic downsizing effect, so the overall evolution of AGN luminosities in \fable \ appears to be realistic.

At low redshifts, observational searches according to our model are likely only uncovering the tip of the iceberg. This issue is explored in more detail in the lower panel where the colour coding shows the average ratio between the (bolometric) BH luminosity $L_\mathrm{BH}$, and the central (bolometric) stellar luminosity $L_\mathrm{stellar}$. Here we take the central stellar luminosity to be the luminosity generated by all of the star particles within the stellar half mass radius. We estimate the luminosity of each star particle using the up-to-date \citet{Bruzual2003} stellar population synthesis models for a \citet{Chabrier2003} IMF, and calculate the bolometric luminosity as a function of stellar age and metallicity.

The majority of AGN in low-mass galaxies are outshone by the stars in their host galaxies - even for highly accreting BHs. Therefore resolved studies as well as multi-wavelength surveys, such as probing the IR and the X-ray regime, are crucial tools for mapping this elusive population (see Section~\ref{sec:intro}).

\subsection{Scaling relations}\label{subsec:scalings}
\begin{figure}
    \centering
    \includegraphics[width=\columnwidth]{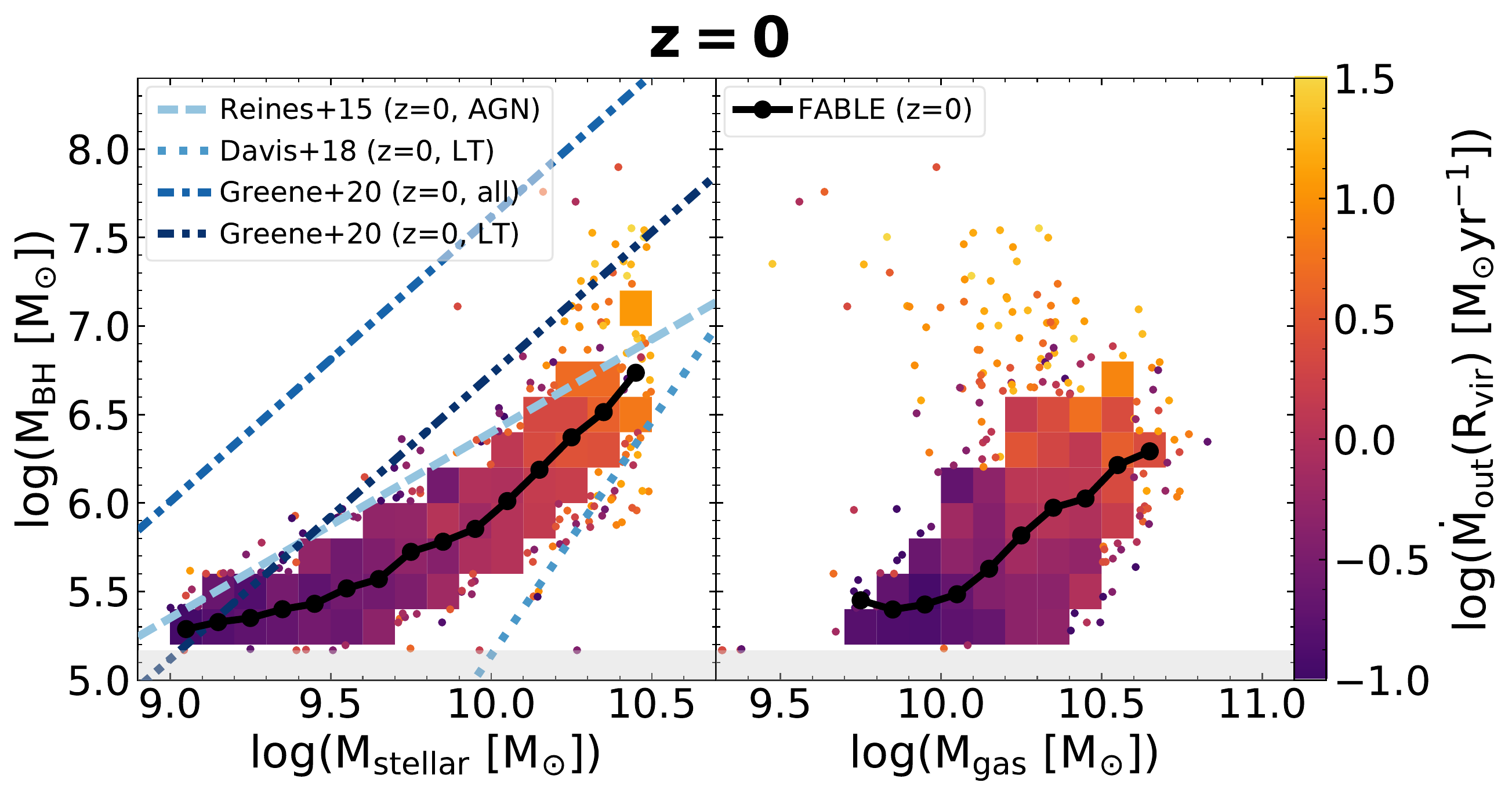}\\
    \includegraphics[width=\columnwidth]{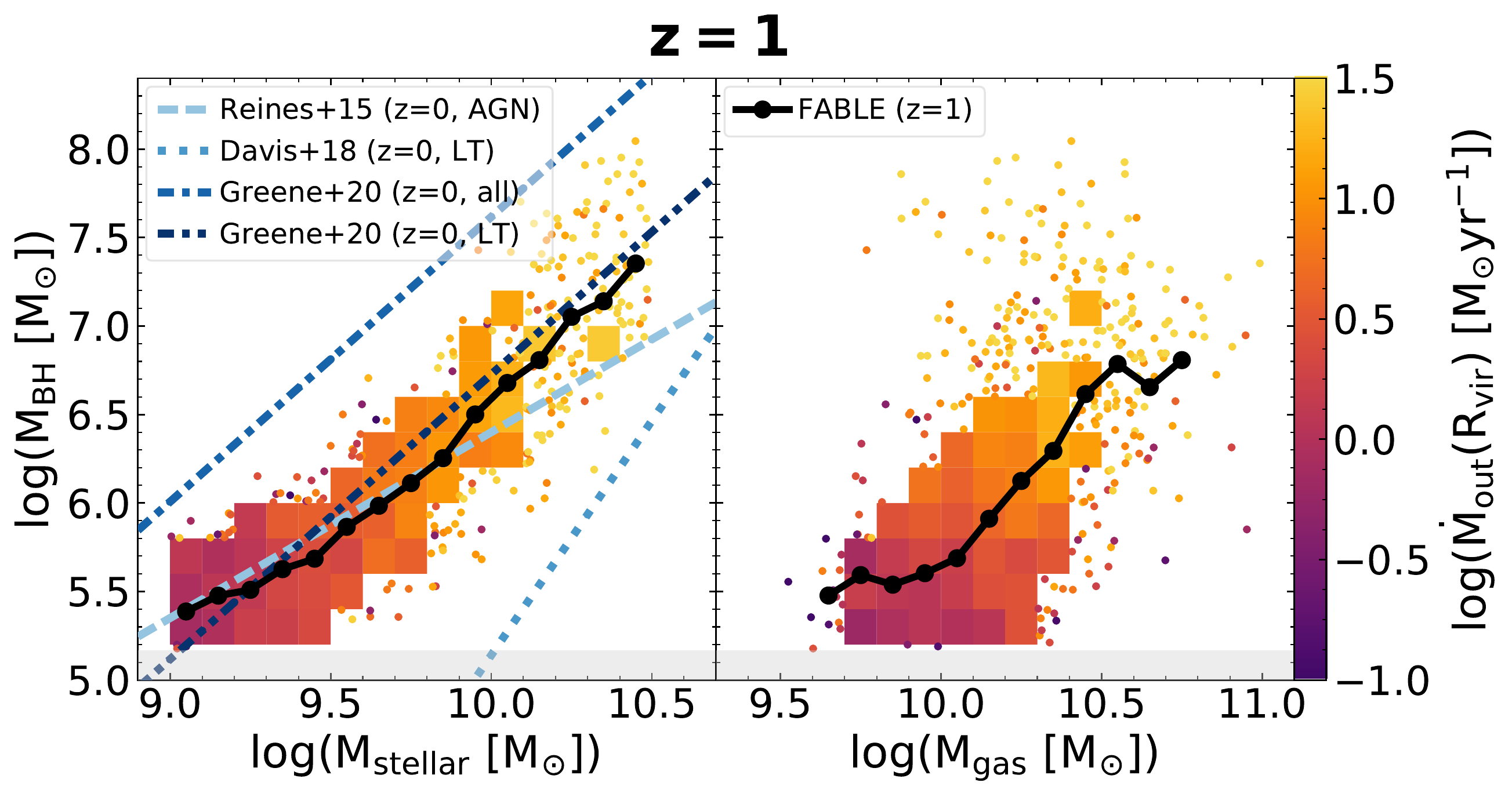}
    \caption{Scaling relations of BH mass, $M_\mathrm{BH}$, against stellar mass, $M_\mathrm{stellar}$, (LHS) and gas mass, $M_\mathrm{gas}$, (RHS), at $z=0$ (top panel) and $z=1$ (bottom panel). The grey-shaded area indicates the seeding floor at $M_\mathrm{BH,seed}= 10^{5} \ h^{-1}\, \Msun$. The \fable \ binned mean scaling relations (bin width 0.1 dex, at least ten objects per bin) are shown as solid black curves with the bin midpoints indicated by filled black circles. Some of the observed $M_\mathrm{BH} - M_\mathrm{stellar}$ scaling relations from the literature, as listed in the legend, are also shown for comparison. These are broadly in agreement with the results from \fable, though the \fable \ BHs are undermassive compared to most of the observed scaling relations. We also show the binned distribution of the whole \fable \ low-mass galaxy sample with colour-coded histograms. 2D bins with at least ten objects are colour-coded by the mean outflow rate of the total gas at $R_\mathrm{vir}$, and otherwise we plot the individual objects colour-coded by their respective outflow rates to indicate the behaviour of outliers. Higher outflow rates are obtained for BHs with larger positive offsets from the $M_\mathrm{BH} - M_\mathrm{gas}$ relation. At $z=1$, the $M_\mathrm{BH}$/$M_\mathrm{stellar}$ and $M_\mathrm{BH}$/$M_\mathrm{gas}$ ratios increase and higher outflow rates occur.}
    \label{fig:ScalingRelations}
\end{figure}
We move on to investigate the relationship between the BHs and the stellar and gaseous components of their host galaxies. Figure~\ref{fig:ScalingRelations} shows the scaling relations of BH mass, $M_\mathrm{BH}$, against stellar mass, $M_\mathrm{stellar}$, (left panels) and BH mass against gas mass, $M_\mathrm{gas}$, (right panels) at redshifts $z=0$ and $z=1$.

Both $M_\mathrm{stellar}$ and $M_\mathrm{gas}$ are calculated within twice the stellar half mass radius, so these are scaling relations for a proxy for the total galaxy mass, rather than comparing just to the bulge component. We use the total galaxy mass since bulge decompositions are rather difficult for observations at the low-mass end \citep[e.g.][]{MacArthur2003,Reines2015,Greene2019} due to the high resolution requirements \citep[although see][]{Schutte2019}.

To investigate whether there is a link between the scatter in the scaling relations and feedback activity in \fable, we define 2D bins (bin dimensions: 0.1 dex $\times$ 0.2 dex) with at least ten objects and colour-code them by their mean outflow rate at $R_\mathrm{vir}$ (using the outflow rate definition from Section~\ref{subsec:OutflowMethods} for the total gas). For more sparsely-populated bins, we plot the individual objects colour-coded by their respective outflow rates. This allows us to show the mean outflow properties of the low-mass galaxy population, whilst also indicating the behaviour of outliers.  

The binned mean scaling relations (bin width 0.1 dex, at least ten objects per bin) for the \fable \ galaxies are shown by the solid black lines with the filled black circles indicating the bin midpoint values. For comparison, we plot some of the observed $M_\mathrm{BH} - M_\mathrm{stellar}$ relations. Here, we mainly focus on relations derived for late-type galaxies (marked by 'LT'). These are most applicable to our low-mass galaxy sample as late-type (star-forming) galaxies dominate the local GSMF at the low-mass end \citep[e.g.][]{Muzzin2013, Vogelsberger2014}. This distinction is important as scaling relations obtained from late-type galaxies have a lower normalization than scaling relations derived from early-type galaxies \citep{Greene2010,Greene2016a,Greene2019, Reines2015,Lasker2016,Davis2018}. 

The relation shown from \citet{Reines2015} is derived from a sample of local AGN (a significant number of which are spirals or discs), mostly containing broad-line AGN but also a subsample of dwarfs and reverberation-mapped AGN. This relation also has a significantly lower normalization than for early-type galaxies \citep[see][for a detailed discussion]{Reines2015}. Moreover, we include the scaling relation derived by \citet{Davis2018} from a sample of 40 spiral galaxies with directly measured BH masses\footnote{As in \citet{Graham2018c}, we use the relation obtained from the symmetric fit using the FITEXY routine to extrapolate the relation to lower masses.}. Finally, we show the relations obtained by \citet{Greene2019}. These are based on the dynamical sample from \citet{Kormendy2013} as well as additional dynamical measurements published since then \citep{Greene2016a, Saglia2016, Krajnovic2018, Thater2019}, particularly at low masses \citep{Brok2015,Nguyen2018,Nguyen2019}, as well as upper limits \citep{Barth2009,Neumayer2012,DeLorenzi2013}. In addition to the relation derived from the late-type \citet{Greene2019} sample, we also show the relation derived from the whole \citet{Greene2019} sample, including both early-type and late-type galaxies. This relation has a greater than $0.7$ dex higher normalization, demonstrating the significant effect of including early-type galaxies in the analysis.

It should be noted that there is a scarcity of dynamical BH mass measurements below $M_\mathrm{stellar} = 3 \times 10^{10}\ \Msun$, so care should be taken when comparing any of the above observed relations with simulated low-mass galaxies. Broad-line AGN can be used as an alternative means to establish a scaling relation for observed low-mass galaxies, however this comes with its own difficulties, as the virial factor $f_\mathrm{vir}$ still carries high uncertainties. 

An additional complication is introduced by the different stellar mass-to-light ratios used in observational studies \citep[see discussion in][]{Reines2015,Davis2018,Davis2018a}. Note that this has no effect on the slope of the scaling relation, just on the normalization. Fully adjusting for this effect would be beyond the scope of this paper, however, we reduce the stellar masses from \citet{Greene2019}, which were obtained from the \citet{Bell2003} fitting functions, by -0.093 dex \citep{Gallazzi2007,Zibetti2009}, so that they are appropriate for a Chabrier IMF, as adopted in \fable. For the \citet{Reines2015} and \citet{Davis2018} relations, we make no adjustments, as these stellar masses were calculated assuming a Chabrier IMF and we do not attempt to harmonise the stellar mass measurements beyond a common IMF. 

Keeping the above caveats in mind, we note that the local BH population in \fable \ is undermassive compared to most of the observed scaling relations, with the exception of the \citet{Davis2018} relation. This relation has a significantly lower normalization than the other observed relations and brackets the lower end of the \fable \ distribution. The closest agreement is reached with the \citet{Reines2015} relation, though the \fable \ mean relation lies consistently below this observed relation with discrepancies from ~ $0.1 - 0.4$ dex. The late-type relation from \citet{Greene2019} brackets the upper end of the \fable \ distribution, whilst the overall relation from \citet{Greene2019}, including both early-type and late-type galaxies, lies significantly above the mean relation derived from \fable \ and has hardly any overlap with the simulation data. 

We also caution that at the low-mass end of the BH mass distribution, where BH masses have not grown considerably from the seed mass of $M_\mathrm{BH,seed}= 10^{5} \ h^{-1}\, \Msun$, alternative seeding models could affect the simulated scaling relations.

Despite the uncertainties regarding the normalization of the observed relations, Figure~\ref{fig:ScalingRelations} suggests that \fable \ may underproduce the masses of AGN in low-mass galaxies at $z=0$ and that any effect we find from AGN on low-mass galaxies in \fable \ might constitute a lower limit of the possible effect of AGN feedback in that mass range. This is also supported by the absence of high-luminosity X-ray dwarfs from \fable \ at low redshifts (see Section \ref{subsec:xrayprops}).

As we move to $z=1$, we find a significantly higher $M_\mathrm{BH}/M_\mathrm{stellar}$ ratio, suggesting that the BHs grew faster than their host galaxies were assembled in the \fable \ simulation, similarly to what has been found in Illustris \citep{Sijacki2015a}. Whether the observed scaling relations evolve at higher redshifts is still controversial. While some studies find that the $M_\mathrm{BH}/M_\mathrm{stellar}$, or equivalently the $M_\mathrm{BH}/L_\mathrm{stellar}$, ratio increases towards higher redshifts \citep[e.g.][]{Kormendy2013,Bongiorno2014,Park2014,Yang2018,Ding2020}, others find that there is no evolution up to $z \sim 1-2.5$ \citep[e.g.][]{Schramm2013,Sun2015,Suh2020}.

However note that, as with the local galaxies, these results are mainly just extrapolated from the high-mass end of the galaxy population. High-redshift studies of BHs in low-mass galaxies are only just at the beginning (see Sections~\ref{sec:intro} and \ref{subsec:xrayprops}). The next generation X-ray missions (\textit{Athena}, \textit{Lynx}) and deep spectroscopic surveys with the \textit{Roman Space Telescope} and JWST will make it possible to obtain significant samples of AGN in low-mass galaxies at intermediate redshifts \citep[see][]{Greene2019}. This will allow us to test our prediction that the $M_\mathrm{BH}/M_\mathrm{stellar}$ ratio should increase with redshift.

Focusing now on the relationship between $M_\mathrm{BH}$ and $M_\mathrm{gas}$, we note that the gas mass scaling relations have a similar slope to the stellar mass scaling relations. Furthermore, the $M_\mathrm{BH}/M_\mathrm{gas}$ ratio likewise increases from $z=0$ to $z=1$. However, the $M_\mathrm{BH}/M_\mathrm{gas}$ ratio has a significantly higher amount of scatter than the quite tightly correlated $M_\mathrm{BH} - M_\mathrm{stellar}$ relation. Interestingly, BHs with large positive offsets from the mean $M_\mathrm{BH} - M_\mathrm{gas}$ relation tend to have higher mass outflow rates, indicating that these overmassive AGN are able to drive more powerful outflows. For the stellar mass relation, the trend is weaker as the outflow activity also increases for higher stellar masses due to increased stellar feedback.

The increased outflow rates for overmassive BHs at a fixed gas mass provide an interesting hint that AGN can enhance galactic outflows once enough gas mass is available. The relationship between overmassive BHs and host galaxy properties is investigated more closely in the next section.

\subsection{Overmassive black holes and galaxy properties} \label{subsec:offsetBHs}
\begin{figure*}
	\includegraphics[width=\textwidth]{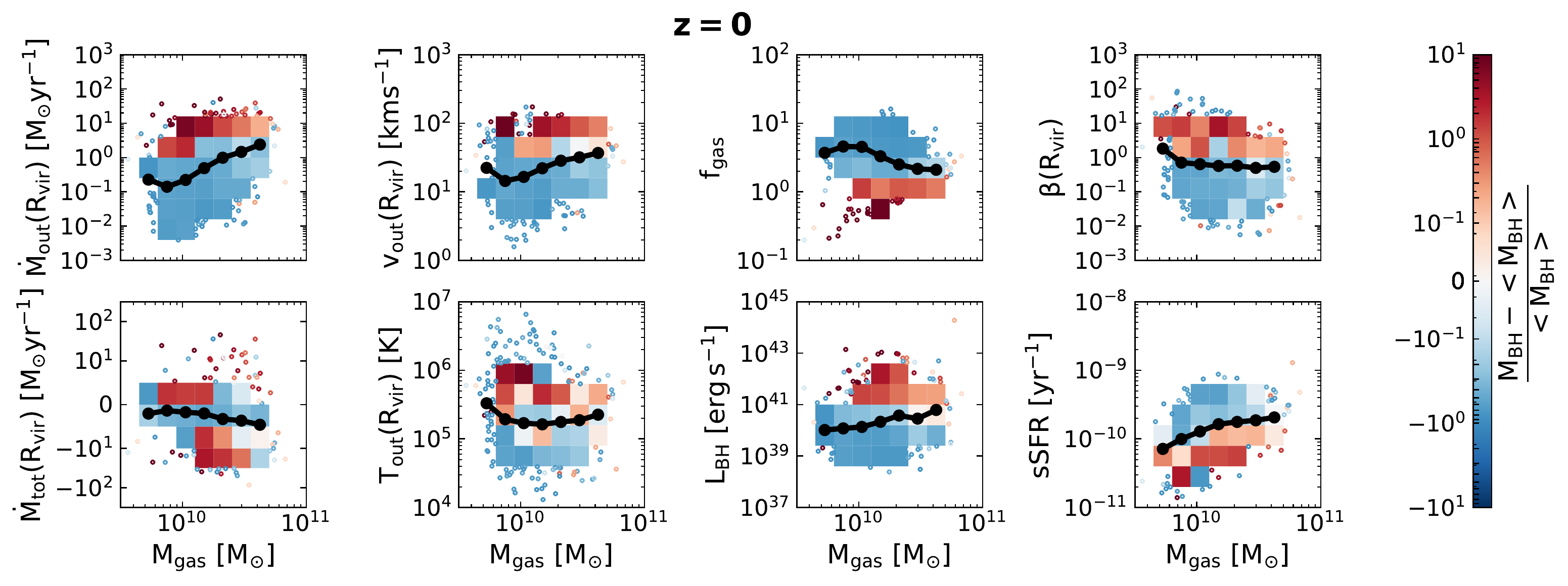}\\
	\includegraphics[width=\textwidth]{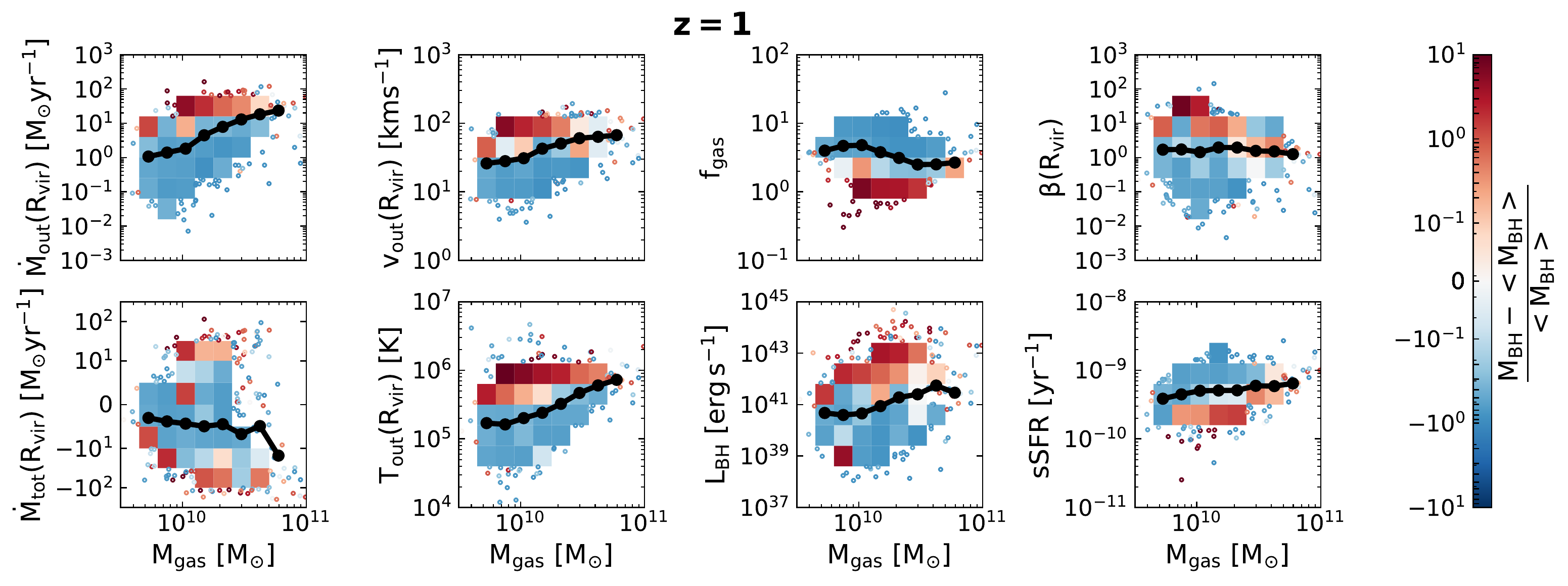}
    \caption{Outflow characteristics of the total gas and galaxy properties against total gas mass at $z=0$ (upper panel) and $z=1$ (lower panel) for the \fable \ low-mass galaxy sample. In each case the binned mean relation is shown as a solid black curve with the filled black circles indicating the bin midpoints. The colour coding of the distribution indicates the offset from the $M_\mathrm{BH} - M_\mathrm{gas}$ scaling relation from Figure~\ref{fig:ScalingRelations}, with blue for undermassive and red for overmassive BHs. 2D bins with at least ten objects are colour-coded according to the mean BH mass offset, and otherwise we plot the individual objects colour-coded by their respective BH mass offsets. Galaxies with overmassive BHs have increased outflow and inflow rates, leading to a bimodal distribution for the total mass flow rate. The outflows are faster and hotter in the overmassive regime. Overmassive BHs are also associated with reduced gas fractions and increased BH luminosities. Furthermore, the mass loading factor $\beta$ is increased, whilst the sSFRs are suppressed.}
    \label{fig:PhysicalPropsBH}
\end{figure*}

\begin{figure}
    \centering
    \includegraphics[width=\columnwidth]{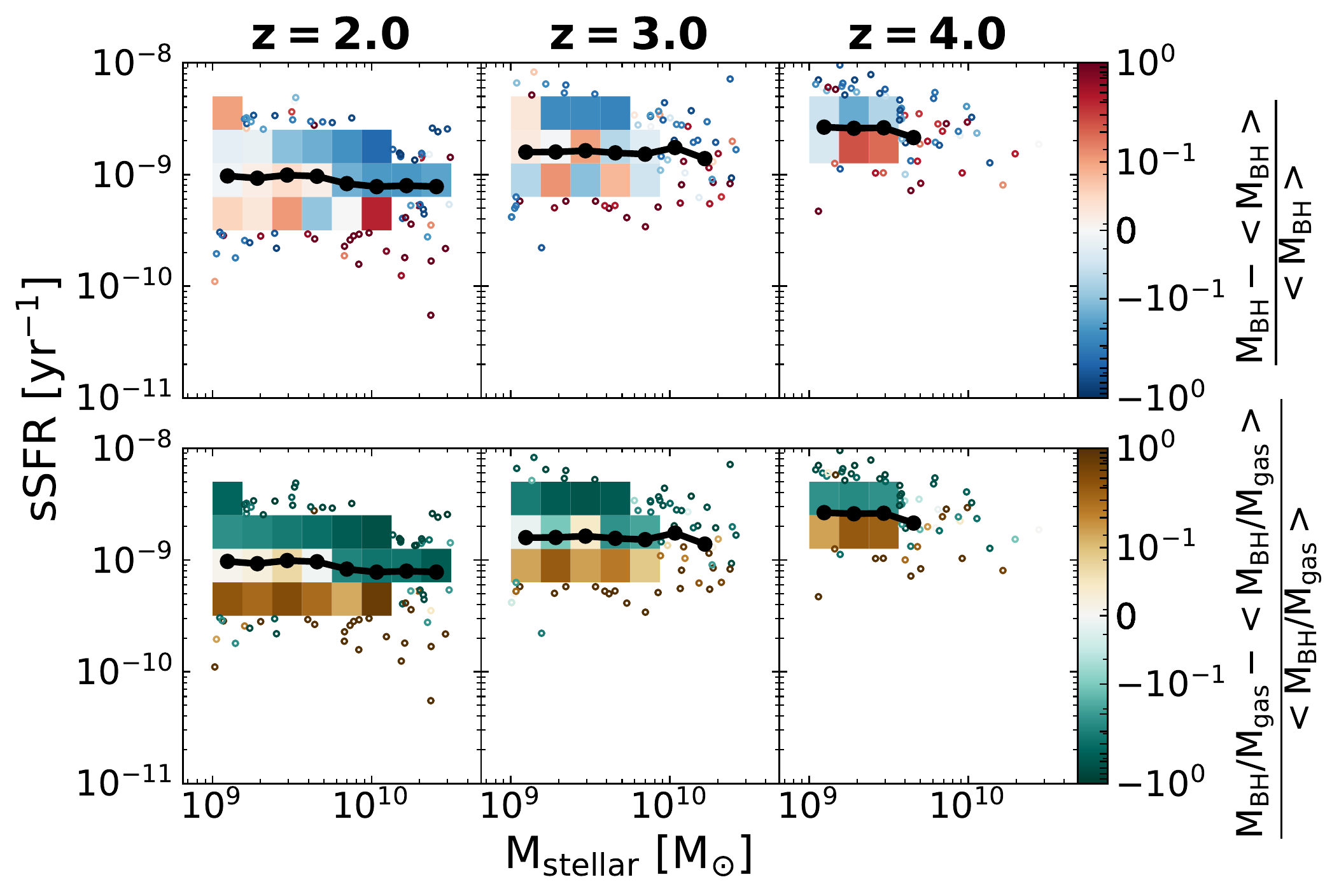}
    \caption{High-redshift evolution of the sSFR against stellar mass. At each redshift, the binned mean relation is shown as a solid black curve with the filled black circles indicating the bin midpoints. In the upper panel, the colour coding indicates the offset from the $M_\mathrm{BH} - M_\mathrm{stellar}$ scaling relation, with red for overmassive BHs and blue for undermassive BHs. In the lower panel, the colour coding indicates the offset from the mean $M_\mathrm{BH}/M_\mathrm{gas}$ ratio of the respective stellar mass bin, with brown for above-average and green for below-average $M_\mathrm{BH}/M_\mathrm{gas}$ ratios. 2D bins with at least ten objects are colour-coded according to the bins' mean offsets, and otherwise we plot the individual objects. At high redshifts, overmassive BHs are correlated with suppressed sSFRs across the whole stellar mass range (contrary to the low-redshift case, see Figure~\ref{fig:PhysicalPropsBHStellar}). The correlation is even stronger when the $M_\mathrm{BH}/M_\mathrm{gas}$ ratio is considered, indicating that at high-redshift, the BHs are able to drive the gas out of dwarf galaxies and suppress star formation.}
    \label{fig:sSFR_overmassive_BHs}
\end{figure}

\begin{figure*}
    \centering
    \includegraphics[width=\columnwidth]{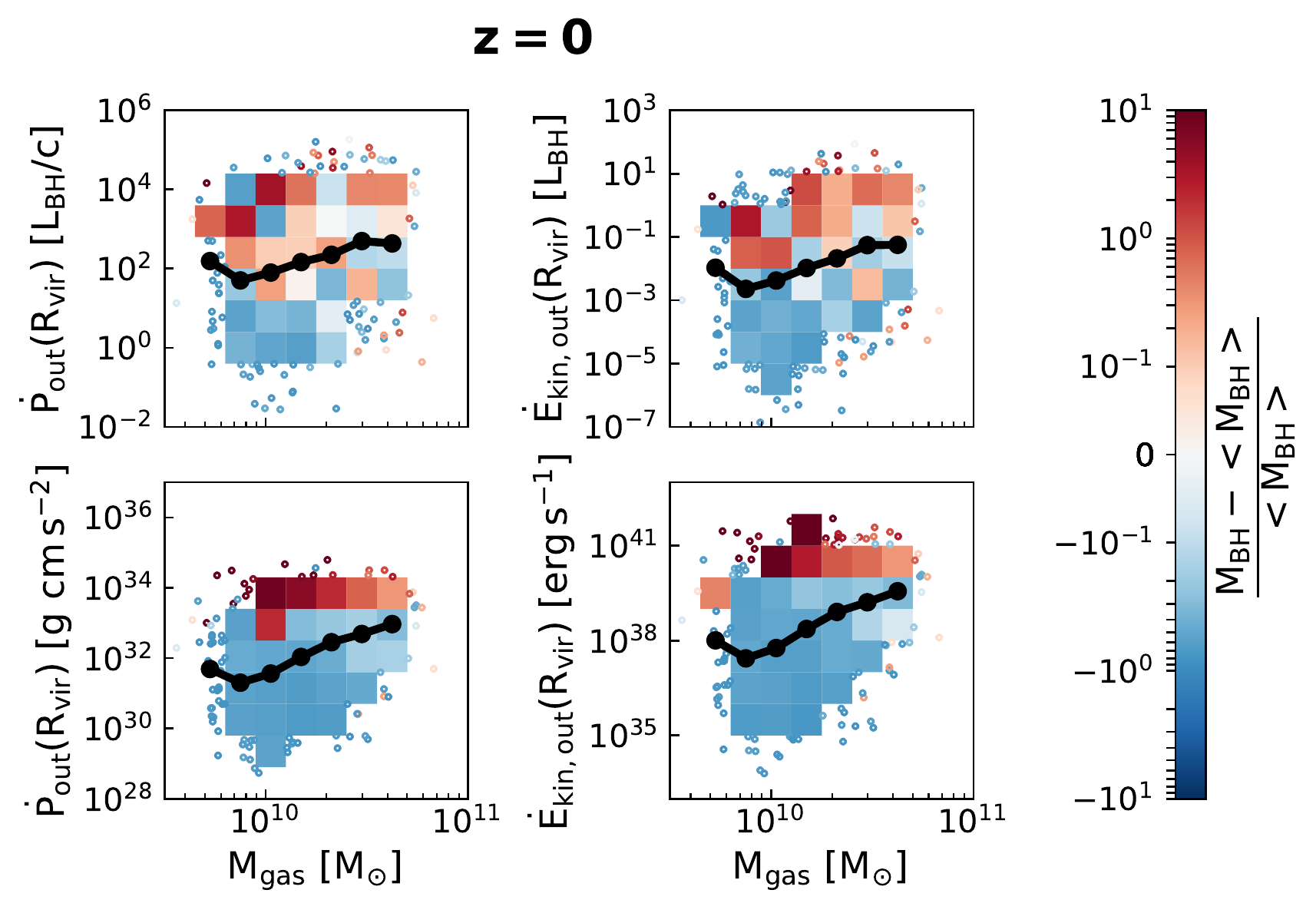}
    \includegraphics[width=\columnwidth]{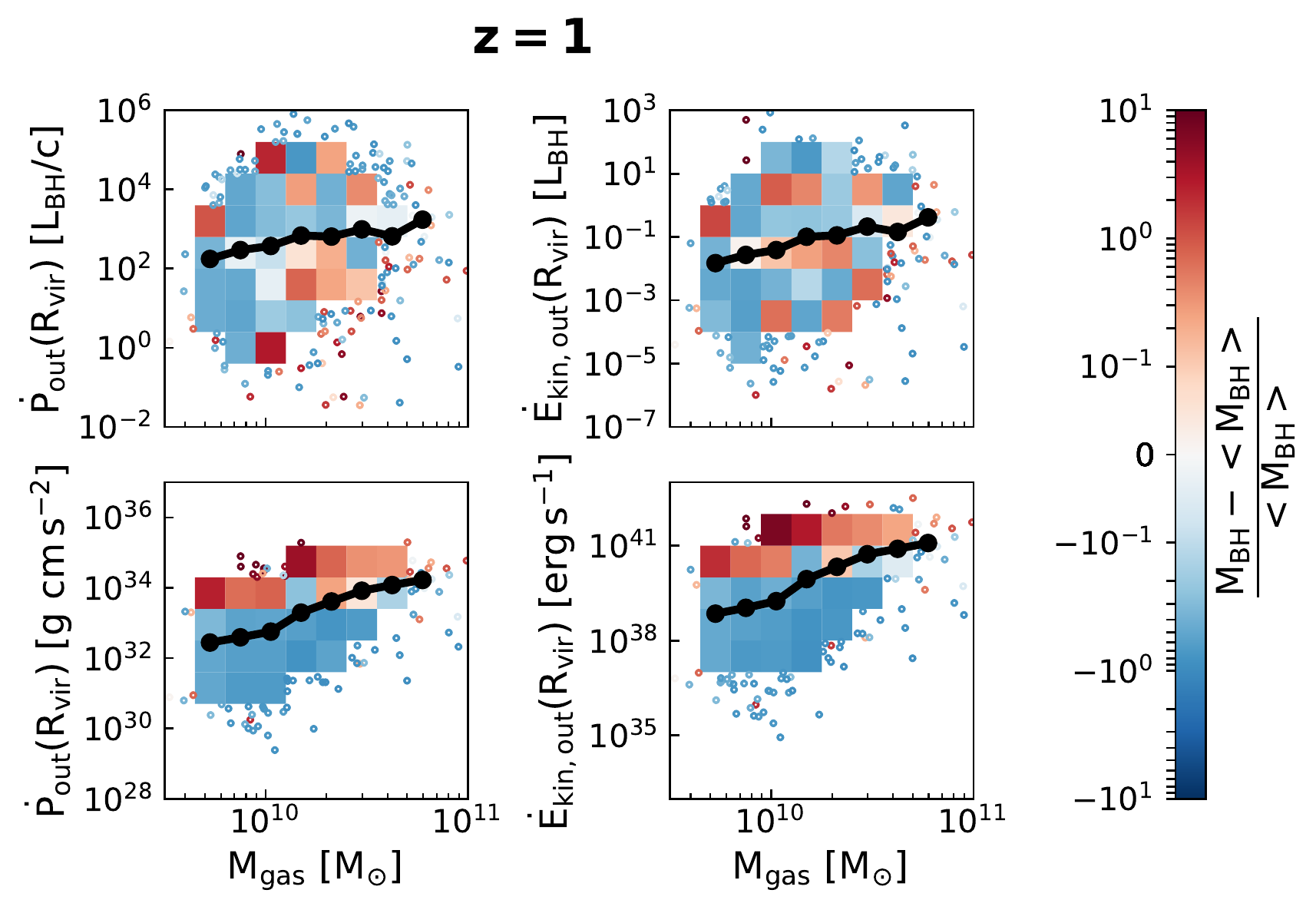}
    \caption{Momentum and kinetic energy outflow rates of the total gas against total gas mass at $z=0$ (left panel) and $z=1$ (right panel) for the \fable \ low-mass galaxy sample. The top row is normalised by the BH luminosity whilst the bottom row shows the outflow quantities in cgs units. The binned mean relations are shown as solid black curves, with the bin midpoints indicated by the filled black circles. Colour coding indicates the offset from the $M_\mathrm{BH} - M_\mathrm{gas}$ scaling relation from Figure~\ref{fig:ScalingRelations}, with blue for undermassive and red for overmassive BHs. Where there are fewer than ten objects per 2D bin, the individual objects are plotted. In cgs units, both momentum and energy outflow rates are clearly enhanced for overmassive BHs. However normalizing by the BH luminosity washes out this correlation at $z=1$ due to strong coupling between BH masses and luminosities (see Figure~\ref{fig:PhysicalPropsBH}).}
    \label{fig:mom_and_egy_rates}
\end{figure*}

Figure~\ref{fig:PhysicalPropsBH} shows various outflow quantities for the total gas at $R_\mathrm{vir}$ (mass outflow rate $\dot{M}_\mathrm{out}$, total mass flow rate $\dot{M}_\mathrm{tot}$, outflow velocity $v_\mathrm{out}$, outflow temperature $T_\mathrm{out}$, mass loading factor $\beta$) as well as integrated galaxy properties (gas mass to stellar mass ratio $f_\mathrm{gas}$, BH luminosity $L_\mathrm{BH}$, specific SFR (sSFR)) plotted against $M_\mathrm{gas}$, both at $z=0$ and at $z=1$. The outflow properties at $R_\mathrm{vir}$ are calculated within a spherical shell of width $\mathrm{\Delta}r = 20 \ \mathrm{kpc}$. For reference, we also checked the outflow properties at $0.2R_\mathrm{vir}$, where we used a shell of width $\mathrm{\Delta}r = 10 \ \mathrm{kpc}$. See Section~\ref{subsec:OutflowMethods} for more details on the calculation of the outflow properties for the total gas.

We include the whole low-mass galaxy sample and colour-code the distribution by the offset from the mean $M_\mathrm{BH} - M_\mathrm{gas}$ relation from Figure~\ref{fig:ScalingRelations}. Overmassive BHs, with BH masses above the mean BH mass <$M_\mathrm{BH}$> for a given gas mass are shown in red, whilst undermassive BHs are shown in blue. As in Figure~\ref{fig:ScalingRelations}, we show binned mean quantities where there are at least ten objects per bin, and otherwise the individual objects are plotted. In each case, we also show the binned mean relation, where there are at least ten objects per 0.15 dex gas mass bin, as a solid black line, with the bin midpoint values shown as filled black circles.

We plot the variation of outflow and galaxy properties with $M_\mathrm{BH}$ at fixed gas mass as the gas supply is a crucial quantity for the effectiveness of the AGN. On the one hand, the AGN needs a sufficient amount of gas to be able to accrete efficiently. On the other hand, AGN feedback can only drive large-scale outflows if there is a sufficient amount of gas to couple to in the host galaxy. To have a fair comparison, it is therefore important to keep $M_\mathrm{gas}$ constant. Figure~\ref{fig:PhysicalPropsBH} shows that there are strong correlations between the offset from the $M_\mathrm{BH} - M_\mathrm{gas}$ relation and all of the quantities considered at fixed gas mass. However, note that we would expect these correlations to be amplified by the tight relation between $M_\mathrm{BH}$ and $M_\mathrm{stellar}$ (see~Figure \ref{fig:ScalingRelations}), i.e. overmassive BHs at fixed gas mass will also tend to be associated with higher stellar masses. To ensure that the trends from Figure~\ref{fig:PhysicalPropsBH} are not just driven by increased stellar feedback activity, we show the same plot for overmassive BHs at fixed stellar mass (see Appendix~\ref{appsec:offsetbhsstellar}). Broadly, the outflow trends are recovered at fixed stellar mass, though the separation of the sample into undermassive and overmassive BHs is less clear-cut. This is partly due to the small scatter in $M_\mathrm{BH}$ at fixed $M_\mathrm{stellar}$. In addition, as discussed above, the AGN's ability to influence its host galaxy is tightly correlated with gas availability. Therefore not keeping the gas mass fixed further blurs the correlation, as there will be a wide range of gas masses for a given stellar mass. We will point out differences between the $M_\mathrm{gas}$-dependent and $M_\mathrm{stellar}$-dependent plots as we discuss the different quantities.

The first column of Figure~\ref{fig:PhysicalPropsBH} shows the mass outflow rates as well as the total mass flow rates. At both redshifts, overmassive BHs are associated with increased outflow as well as increased inflow rates. For the total mass flow rate this leads to a bimodal distribution with both extreme inflow and outflow rates correlated with overmassive BHs. This bimodality is to be expected as on the one hand large gas inflow rates will allow BHs to grow more quickly and become overmassive, and on the other hand these overmassive BHs are then able to drive more powerful outflows than their undermassive counterparts. Outflow rates are higher at $z=1$ than at $z=0$, and in both cases increase towards higher gas masses. At $z=0$, overmassive BHs reach outflow rates up to $\sim 50 \ \Msun \, \mathrm{yr^{-1}}$, whilst at $z=1$ these have outflow rates of up to $\sim 160 \ \Msun \, \mathrm{yr^{-1}}$. Note that we obtain the same trends at fixed stellar mass, however the overmassive BH population is not as clearly separated and the variations in BH masses at fixed $M_\mathrm{stellar}$ are smaller.

The second column shows the outflow velocity and temperature of the total gas at $R_\mathrm{vir}$. Outflows are on average faster and reach higher temperatures for overmassive BHs. Furthermore, the outflow velocities and temperatures are on average higher at $z=1$ than at $z=0$. This is particularly noteworthy considering that, at fixed halo mass, the average gas density within $R_\mathrm{vir}$ will be higher at higher redshifts.

We note that when we inspect the same outflow properties at $0.2R_\mathrm{vir}$, whilst the trends for the mass flow rates remain the same, the trends for $v_\mathrm{out}$ and $T_\mathrm{out}$ with BH mass offset are much more distinct. This suggests that the AGN mainly accelerates and heats the gas at a local level, while for the large-scale outflows the difference is somewhat less significant. This is especially true for the distribution at fixed $M_\mathrm{stellar}$ where there is not a strong link between outflow temperatures and BH mass offsets at $R_\mathrm{vir}$, however, there is a clear correlation at $0.2 R_\mathrm{vir}$.

At $z=0$, the mean outflow temperature stays relatively constant with increasing gas mass at $\sim 2\times 10^{5} \ \mathrm{K}$. At $z=1$, the outflow temperature increases with gas mass from $\sim 2 \times 10^{5}$ to $\sim 7 \times 10^{5} \ \mathrm{K}$. The outliers at both the low-temperature and the high-temperature end are dominated by undermassive BHs, suggesting that the link between BH mass and outflow temperature at $R_\mathrm{vir}$ is slightly weaker. We found that the high-temperature outliers at $0.2 R_\mathrm{vir}$, on the other hand, are overmassive giving further support to the hypothesis that the AGN heating mainly affects the small-scale outflows.

For the $z=0$ galaxies, the mean outflow velocity ranges from $\sim 10 - 40 \ \mathrm{km \, s^{-1}}$ and for the $z=1$ galaxies from $\sim 20 - 70 \ \mathrm{km \, s^{-1}}$. Galaxies with overmassive BHs at both redshifts reach typical outflow velocities of $\sim 100 \ \mathrm{km \, s^{-1}}$, with the outliers reaching up to $\sim 200 \ \mathrm{km \, s^{-1}}$.

Using the virial velocity $V_\mathrm{vir}$ as a proxy for the escape velocity (as in Section ~\ref{subsec:population}), we find that at $z=0$, the mean escape velocity increases with gas mass from $\sim 80 - 140 \ \mathrm{km \, s^{-1}}$, and at $z=1$ from $\sim 100 - 220 \ \mathrm{km \, s^{-1}}$. Therefore, at both redshifts, the fastest outflows should be able to escape the halo, whilst for the systems that lie on the mean relation, the outflow velocities are not high enough to escape the halo. Though note that, as discussed in Section~\ref{subsec:population}, the outflows are multiphase and that even if the outflow velocity for the total gas, as considered here, is too slow to escape the halo, the hot and fast outflow phase may still be able to escape. 

At $0.2 R_\mathrm{vir}$, the outflow velocities are significantly higher with the mean outflow velocity ranging from $\sim 20 - 90\ \mathrm{km \, s^{-1}}$ at $z=0$ and from $\sim 70 - 140\ \mathrm{km \, s^{-1}}$ at $z=1$, implying that the outflows are decelerating. At both redshifts, typical outflow velocities for overmassive BHs are $\sim 200 - 300 \ \mathrm{km \, s^{-1}}$. The properties of AGN-driven outflows in the low-mass regime are still relatively unconstrained by observations due to the high resolution requirements. \citet{Manzano-King2019} presented the first direct detection and measurements of AGN-driven outflows in dwarf galaxies. They find outflow velocities from $375 - 1090 \ \mathrm{kms^{-1}}$ for galaxies with masses $\sim 4\times 10^{8} - 9\times 10^{9} \ \Msun$. However, they measure these outflows in a relatively small region ($1.5 - 3.0 \ \mathrm{kpc}$ from the centre), whilst our outflow measurements at $0.2 R_\mathrm{vir}$ correspond to length scales of $\sim 16 - 44 \ \mathrm{kpc}$ in the dwarf regime. As we find that the \fable \ outflows are decelerating, velocities in the above range could be plausible for the total gas in the central region. Though we cannot test this due to the resolution of \fable, as the gravitational softening length is $\sim 3.5 \ \mathrm{kpc}$ so the galaxies' central regions are not resolved in our simulation.

Next, we investigate the relation between overmassive BHs and gas fractions as well BH luminosities in the third column of Figure~\ref{fig:PhysicalPropsBH}. $f_\mathrm{gas}$ is reduced for large offsets from the mean BH mass, with overmassive BHs mostly found in galaxies with gas fractions $f_\mathrm{gas} \leq 1$. This is in agreement with the observed suppressed gas fractions of low-mass galaxies with AGN activity \citep{Bradford2018, Ellison2019}. Furthermore, the overmassive BHs are also correlated with high BH luminosities suggesting that these overmassive BHs are still actively accreting.

Finally, we investigate the properties related to star formation. The last column shows the mass loading factor at the virial radius $\beta (R_\mathrm{vir})=\dot{M}_\mathrm{out}(R_\mathrm{vir})/\mathrm{SFR}(\leq R_\mathrm{vir})$ and the $\mathrm{sSFR} = \mathrm{SFR}/M_\mathrm{stellar}$ (the latter is evaluated within twice the stellar half mass radius). Note that at $z=0$, 0.5 per cent of objects have a sSFR below $10^{-11}$ (for 0.3 per cent of objects the sSFR is zero). These objects are also extremely gas-depleted and are not plotted here. Overmassive BHs are associated with higher mass loading factors and suppressed sSFRs for the whole gas mass range. For overmassive BHs at $z=0$, we obtain sSFRs in the range $\sim 10^{-11} - 2 \times 10^{10} \ \mathrm{yr^{-1}}$. Whilst at $z=1$, where the mean sSFR is significantly higher, the majority of overmassive BHs have sSFRs in the range $10^{-10} - 6 \times 10^{-10} \ \mathrm{yr^{-1}}$. 

However, there are clear differences when we plot these two (star-formation related) quantities at fixed stellar mass instead (see Figure~\ref{fig:PhysicalPropsBHStellar}). For both $\beta$ and the sSFR there is a turnover in the massive dwarf regime at $\sim 6 \times 10^{9} \ \Msun$ at $z=0$ and $z=1$. Below this mass, the trends from Figure~\ref{fig:PhysicalPropsBH} are either reversed or washed out. This suggests that star formation in \fable \ dwarfs is not affected by AGN feedback at low redshifts ($z < 2$).

For $z\geq2$, however, as the AGN reach higher luminosities (see Figure~\ref{fig:PopulationStatistics}), there is a clear correlation between overmassive BHs and suppressed sSFRs across the whole stellar mass range. This is illustrated in Figure~\ref{fig:sSFR_overmassive_BHs}, where we plot the sSFR against stellar mass for $z=2,3$ and 4. We show the distribution of the sSFRs colour-coded by both the offset from the mean $M_\mathrm{BH} - M_\mathrm{stellar}$ relation (upper panel) as well as by the offset from the mean $M_\mathrm{BH}/M_\mathrm{gas}$ ratio for the respective stellar mass bin (lower panel). At high redshifts, overmassive BHs at fixed $M_\mathrm{stellar}$ are associated with suppressed sSFRs down to the dwarf regime - contrary to the low-redshift case. The correlation is even stronger when considering the offset from the mean $M_\mathrm{BH}/M_\mathrm{gas}$ ratio. This implies that at $z\gtrsim 2$, the AGN feedback can drive the gas out of dwarf galaxies and help regulate star formation, whilst at low redshifts, stellar feedback is the dominant quenching process. However, we note that based on the undermassive \fable \ BHs at $z=0$ compared to the observed scaling relations (see Figure~\ref{fig:ScalingRelations}) and the lack of high-luminosity X-ray in \fable \ (see Figure~\ref{fig:AGNFraction}), this quenching transition at $z\sim2$ could also be delayed to lower redshifts with more efficient BH growth at late times. 

In observations, low-mass galaxies (with a typical stellar mass of $M_\mathrm{stellar} \sim 9\times 10^{9} \Msun$) with overmassive BHs (with respect to the $M_\mathrm{BH} - \sigma$ relation) have been found to also have reduced star formation - although this correlation is less significant than for massive galaxies \citep{Martin-Navarro2018}. This trend is also found in Illustris, across the whole galaxy mass range, whilst for IllustrisTNG the correlation is much weaker \citep{Li2020}. Furthermore, \citet{Sharma2020} find that isolated dwarf galaxies with overmassive BHs in \textsc{Romulus25} experience star formation suppression starting at around $z=2$.

Nevertheless, the enhancement of outflows by AGN is observed throughout cosmic time. We investigate the outflow properties more closely in Figure~\ref{fig:mom_and_egy_rates} where we plot the momentum and kinetic energy outflow rates for the total gas, as defined in Section~\ref{subsec:OutflowMethods}, against gas mass. In the top row, we show these rates in units of the BH luminosity and in the bottom row in cgs units. We use the same colour-coding as in Figure~\ref{fig:PhysicalPropsBH} with red indicating overmassive BHs and blue indicating undermassive BHs with respect to the mean $M_\mathrm{BH} - M_\mathrm{gas}$ relation. 

Overmassive BHs have both higher momentum and higher energy outflow rates. However, once we normalise these rates by the BH luminosity the trends become weaker, as overmassive BHs also have higher BH luminosities (see Figure~\ref{fig:PhysicalPropsBH}). This is particularly true for $z=1$ where the BHs reach significantly higher luminosities and the BH offsets are correlated with the luminosities across a wider gas mass range.

We also investigated the momentum and energy outflow rates plotted against stellar mass (see Figure~\ref{fig:mom_and_egy_rate_stellar} in Appendix~\ref{appsec:offsetbhsstellar}). As with the previous plot, we recover similar trends at fixed stellar mass but the trends are weaker due to the variation in gas supply as well as the small scatter in BH masses at fixed $M_\mathrm{stellar}$. When we normalise the trends at fixed stellar mass by the BH luminosity, the correlation is reversed. This is because the correlation with BH luminosity at fixed stellar mass is much stronger (see Figure~\ref{fig:PhysicalPropsBHStellar}) and therefore able to invert the outflow trends.

\subsection{Kinematic properties: mock MaNGA observations}
\label{subsec:MockMaNGAObs}

In Section~\ref{subsec:offsetBHs}, we demonstrated that overmassive BHs drive more powerful outflows reaching higher velocities, temperatures, and mass loading factors. Next, we assess the potential impact of overmassive BHs on observations of low-mass galaxies.

\citet{Penny2018a} find that dwarf galaxies (which they define as galaxies with $M_\mathrm{stellar} \lesssim 5 \times 10^{9} \ \Msun$) with AGN signatures are more likely to have an ionized gas component with large kinematic offsets with respect to the stellar component (difference between global kinematic PAs is $30 \degree \leq \Delta \mathrm{PA} < 150 \degree$). They interpret this as indirect evidence for AGN-driven outflows which may be responsible for the kinematically misaligned ionized gas.

Here we test this hypothesis by producing mock MaNGA l.o.s. velocity maps for both the stars and the ionized gas following the procedure described in Section~\ref{subsec:MaNGAMethods}. We create these maps at three different redshifts ($z=0.0,0.2,0.4$) with a pixel size of 0.5 arcsec (size of the MaNGA square spaxels) and we convolve the maps with a Gaussian filter of 2.5 arcsec, matching the MaNGA PSF. Further, we only include stars and gas within a 2D aperture with radius $1.5r_\mathrm{eff}$, which corresponds to the spatial coverage of the primary MaNGA sample. Note that \citet{Henden2019} find that \fable \ galaxies with $M_\mathrm{stellar}<10^{11} \  \Msun$ have larger half-mass radii $r_\mathrm{eff}$ than observed galaxies \citep[e.g.][]{Baldry2012} by roughly a factor of two. This means that we can better resolve \fable \ galaxies with the MaNGA PSF than may be the case for the observed galaxies \citep[see Figure A1 in][for the median relationship between galaxy size and stellar mass
for galaxies in the \fable \ simulations compared to the sizes from other simulations as well as observed sizes]{Henden2019}.

To check for the effect of orientation, we create the l.o.s. velocity projections for each galaxy at two fixed orientations: edge-on (inclination angle $\theta =0 \degree$) and inclined by $\theta =45 \degree$. We rotate the galaxies automatically by aligning $\mathbf{L}_\mathrm{stellar}$ with the vertical axis. Note that we only include galaxies that are rotationally supported. See Section~\ref{subsec:MaNGAMethods} for more details on the mock l.o.s. velocity maps.

To assess the impact of the AGN-boosted outflows in galaxies with overmassive BHs, we select the galaxies below the $20^\mathrm{th}$ percentile ($\mathrm{P}_{20}$) and above the $80^\mathrm{th}$ percentile ($\mathrm{P}_{80}$) of the $M_\mathrm{BH}/M_\mathrm{gas}$ distribution for each of the three stellar mass bins (dwarfs with $9.0 \leq \log(M_\mathrm{stellar} \ [\Msun]) < 9.5$, massive dwarfs with $9.5 \leq \log(M_\mathrm{stellar} \ [\Msun]) < 10.0$, and $\mathcal{M^{*}}$ galaxies with $10.0 \leq \log(M_\mathrm{stellar} \ [\Msun]) < 10.5$) at each redshift. Note that we only consider galaxies that are resolved by at least 50 star particles and 50 ionized gas cells within the $1.5r_\mathrm{eff}$ MaNGA aperture (see Section~\ref{subsec:MaNGAMethods} for details). We then create mock MaNGA maps for these galaxies and determine the kinematic offsets between the ionized gas and the stars to assess whether heightened AGN activity produces a significantly higher percentage of kinematically misaligned galaxies.  

\begin{figure*}
    \centering
    \includegraphics[width=\textwidth]{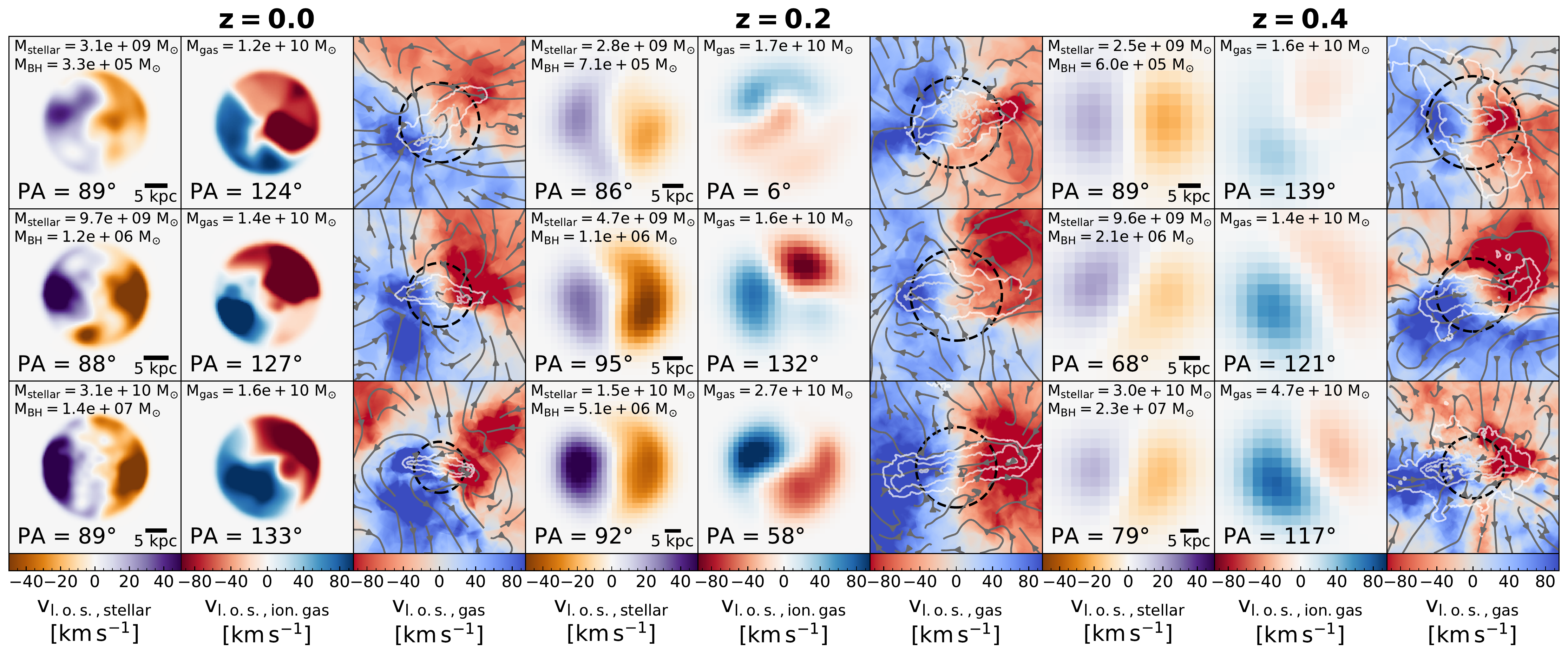}
    \caption{Example mock MaNGA maps of nine different kinematically misaligned \fable \ galaxies (difference in global kinematic PAs between the stellar and ionized gas component lies in the range $ 30 \degree \leq \Delta \mathrm{PA} < 150 \degree$) at $z=0.0$ (left panel), $z=0.2$ (middle panel), and $z=0.4$ (right panel). For each panel, the first and second columns show l.o.s. velocity maps for the stars and ionized gas, respectively, matching the MaNGA resolution and including all resolution elements of the FoF group within the $1.5 r_\mathrm{eff}$ 2D MaNGA aperture. The third column shows the l.o.s. velocity of the total gas at the original resolution of the simulation (projection dimensions: $0.2 R_\mathrm{vir} \times 0.2 R_\mathrm{vir} \times 2R_\mathrm{vir}$). For reference, the $1.5r_\mathrm{eff}$ 2D MaNGA aperture is shown as a dashed black circle, the velocity streamlines are shown in dark grey and the gas density contours are plotted in light grey. We show examples from all of the three stellar mass bins considered: dwarf galaxies ($9.0 \leq \log(M_\mathrm{stellar} \ [\Msun]) < 9.5$, first row), massive dwarf galaxies ($9.5 \leq \log(M_\mathrm{stellar}\ [\Msun]) < 10.0$, second row), and $\mathcal{M^{*}}$ galaxies ($10.0 \leq \log(M_\mathrm{stellar}\ [\Msun]) < 10.5$, third row). All of the example galaxies shown here have a $M_\mathrm{BH}/M_\mathrm{gas}$ ratio above the 80th percentile for their respective stellar mass bin and have been rotated to be viewed as edge-on. In some cases the kinematic misalignment is caused by fast outflows, however other factors, such as inflows or mergers, also play a role.}
    \label{fig:MaNGAMaps}
\end{figure*}

\begin{figure*}
    \centering
    \includegraphics[width=0.33\textwidth]{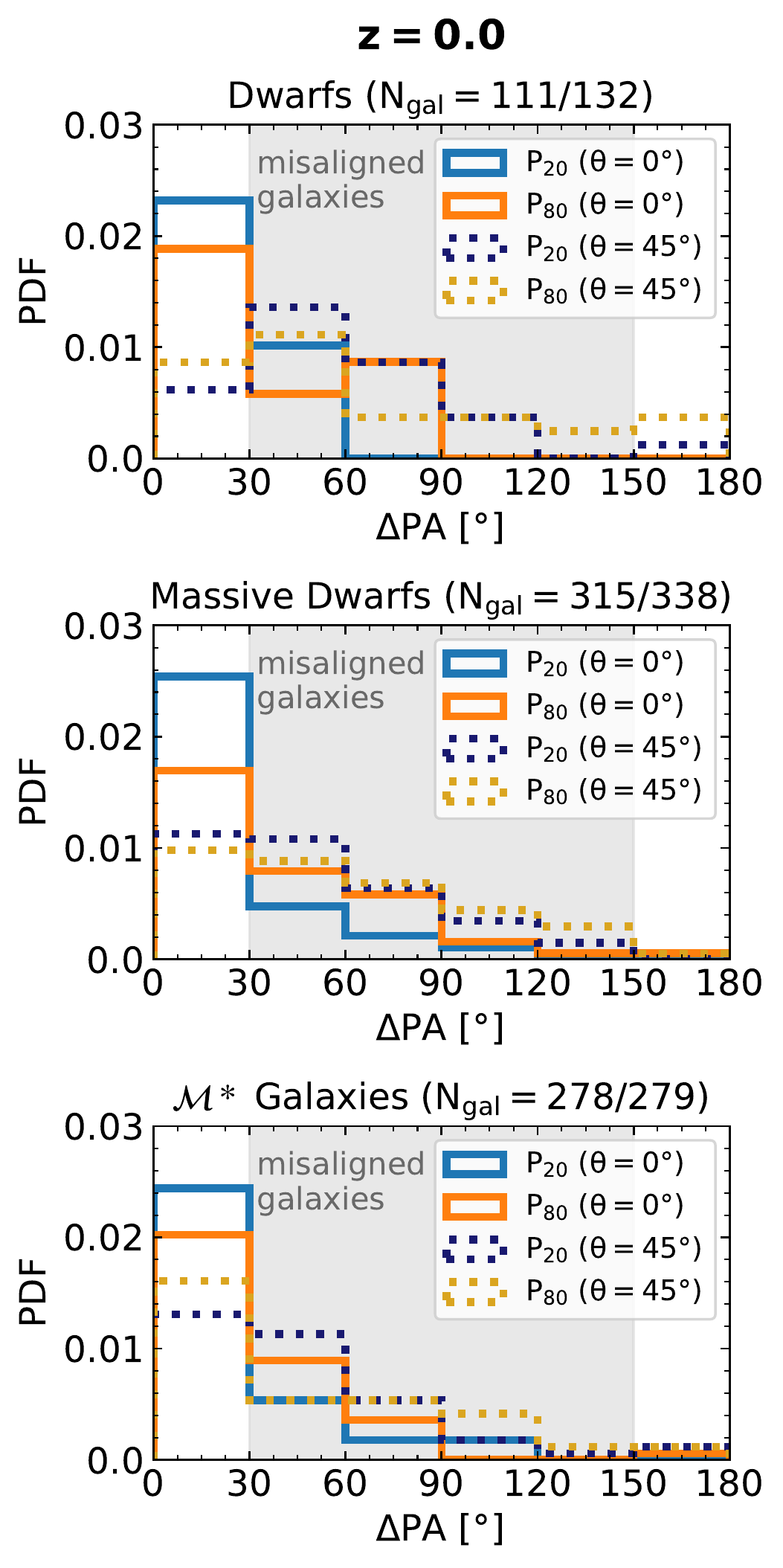}
    \includegraphics[width=0.33\textwidth]{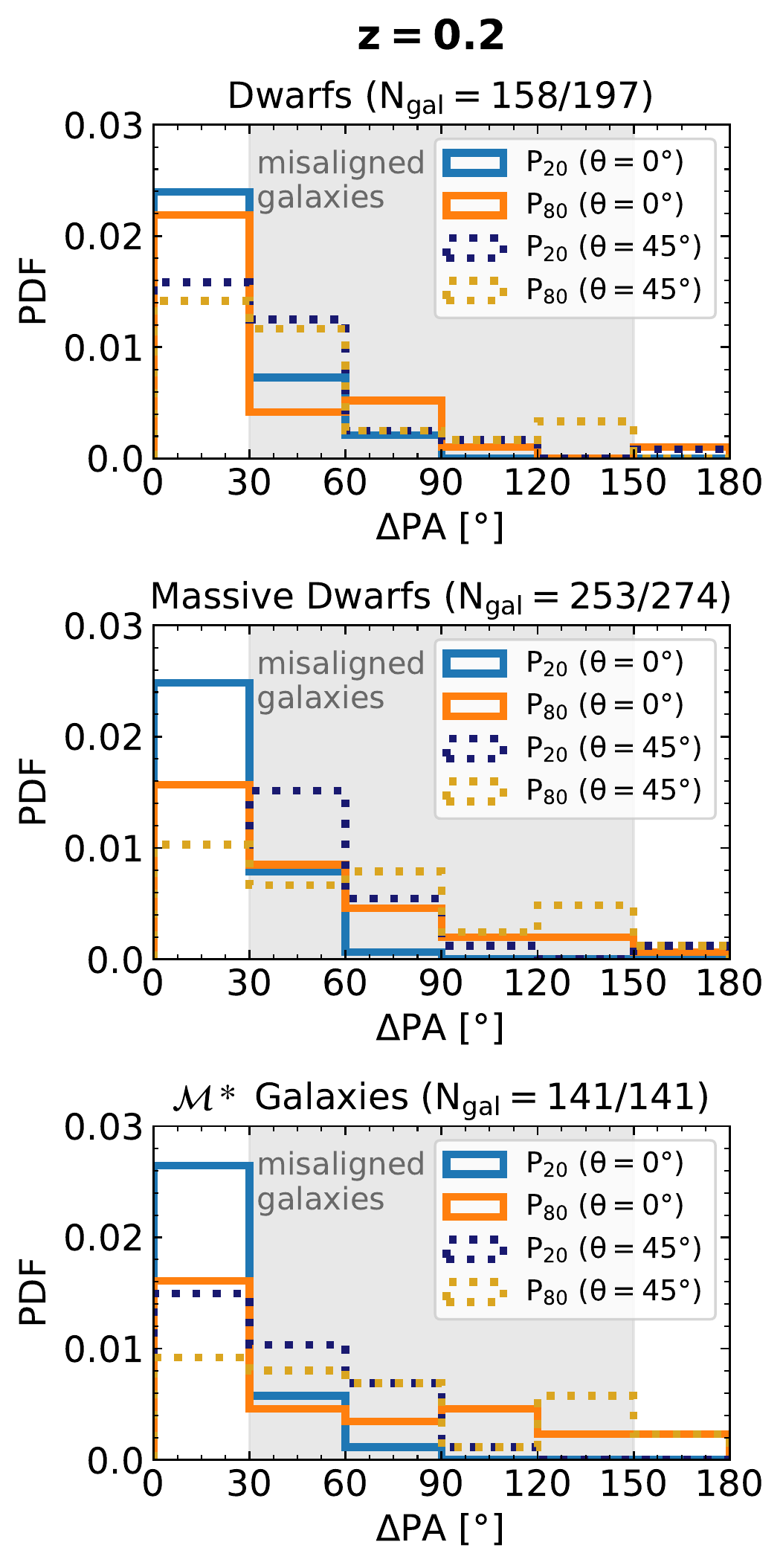}
    \includegraphics[width=0.33\textwidth]{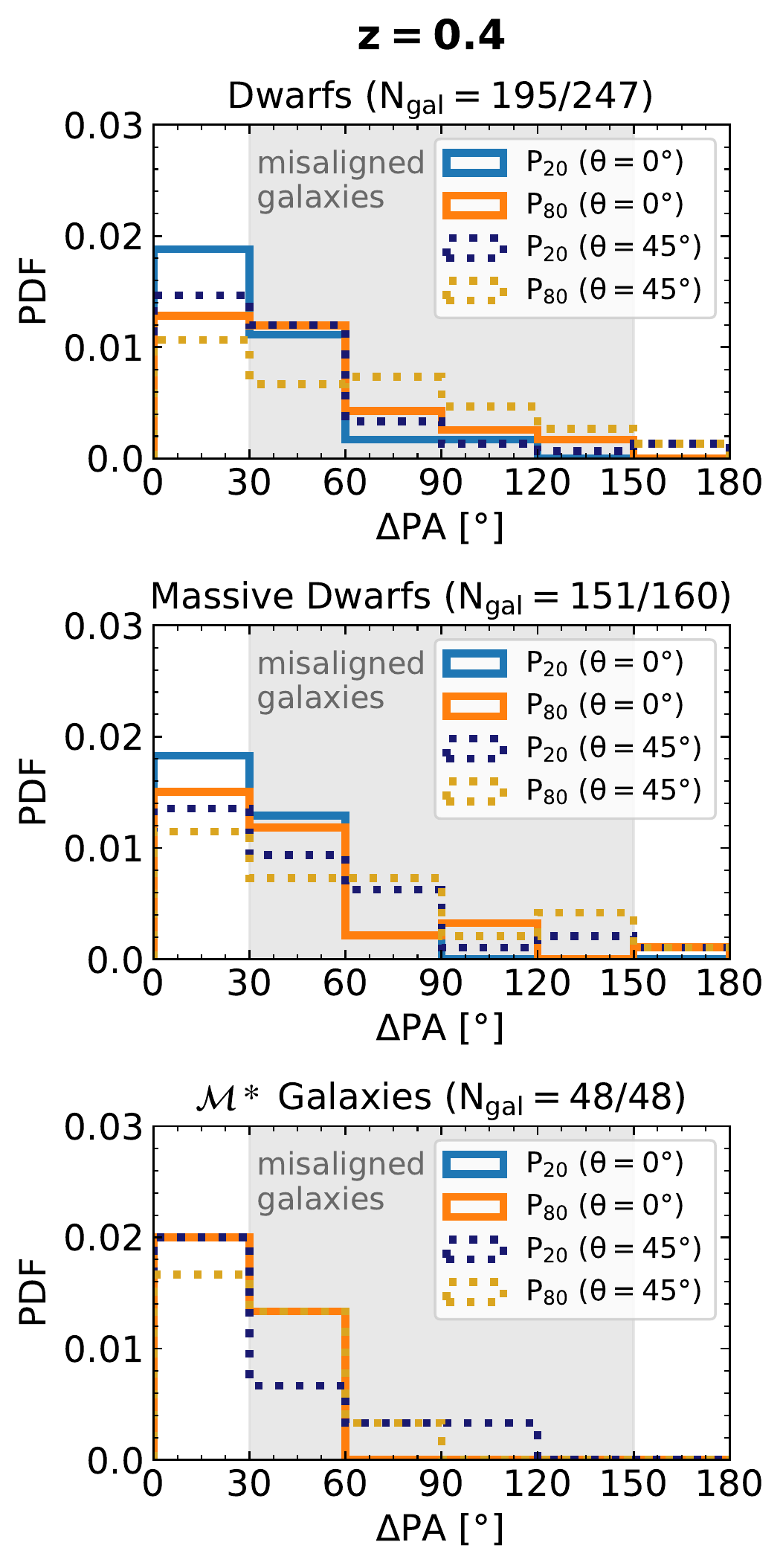}
    \caption{PDFs of the difference between global kinematic PAs of stars and ionized gas, $\Delta \mathrm{PA}$, for $z=0.0$ (left panel), $z=0.2$ (middle panel), and $z=0.4$ (right panel), separated into three mass bins of equal widths in log space: dwarf galaxies ($9.0 \leq \log(M_\mathrm{stellar} \ [\Msun]) < 9.5$, first row), massive dwarfs ($9.5 \leq \log(M_\mathrm{stellar}\ [\Msun]) < 10.0$, second row), and $\mathcal{M^{*}}$ galaxies ($10.0 \leq \log(M_\mathrm{stellar}\ [\Msun]) < 10.5$, third row). Within each mass bin, we select the galaxies that fall below the $\mathrm{20^{th}}$ percentile or above the $\mathrm{80^{th}}$ percentile of the $M_\mathrm{BH}/M_\mathrm{gas}$ ratio distribution. We show the PDFs of $\mathrm{\Delta}$PA for two fixed inclination angles: $\theta= 0 \degree$ (edge-on) and $\theta=45 \degree$. The plot titles give the sizes of the samples of galaxies that are sufficiently resolved and sufficiently rotationally supported for projections at $\theta=0\degree$ and $\theta = 45 \degree$, respectively. Galaxies with overmassive BHs are more likely to be offset across mass bins and redshifts. This difference is even more pronounced for $\theta= 0 \degree$, as rotational features are stronger and therefore only fast inflows or outflows can offset the rotational motion in projection.}
    \label{fig:PADistribution}
\end{figure*}

Figure~\ref{fig:MaNGAMaps} shows example mock MaNGA maps for \fable \ galaxies from each of the three mass stellar bins at $z=0.0,0.2,0.4$, demonstrating some of the qualitative features of interest. Note that all of the projections shown here are at an inclination of $\theta =0 \degree$ (edge-on). All of the example galaxies have overmassive BHs and would be classified as kinematically misaligned. In addition to the mock MaNGA maps, we also show the l.o.s. velocity maps of the whole gas at the original resolution of the simulation (projection dimensions: $0.2 R_\mathrm{vir} \times 0.2 R_\mathrm{vir} \times 2R_\mathrm{vir}$). The $1.5 r_\mathrm{eff}$ MaNGA aperture is indicated by a dashed black circle. Furthermore, we show the gas density contours in light grey and the velocity streamlines in dark grey to highlight the inflowing and outflowing streams around the gas disc.

The kinematic PA is measured from the vertical axis and traces the position of the maximum change in velocity. We determine the kinematic PA from our mock MaNGA maps for both ionized gas and stars using the \texttt{fit\_kinematic\_pa} routine \citep[see][]{Krajnovic2006}. The stellar l.o.s. velocity maps are dominated by the rotational motion of the disc. Since we have aligned $\mathbf{L}_\mathrm{stellar}$ with the vertical axis, we would therefore expect the kinematic PA of the stellar component to be $\mathrm{PA} \sim 90 \degree$. 

For the example galaxies shown here, this is recovered within $\pm 22 \degree$. For the ionized gas component, however, we obtain large kinematic offsets, with respect to the stellar component, from $\Delta \mathrm{PA} = 34 \degree$ to $\Delta \mathrm{PA} = 80 \degree$. This means that the method for determining kinematic PAs is sufficiently robust and the large kinematic offsets between stars and ionized gas are real features. 

The gas kinematics at the original resolution of the simulation (third column of each panel) reveal complex inflow-outflow structures. Fast outflows can cause kinematic misalignment, see e.g. the example $\mathcal{M^{*}}$ galaxy at $z=0.0$ (Figure~\ref{fig:MaNGAMaps}, left panel, third row). However, we identified a number of different physical processes other than outflows that can also result in kinematic misalignment. These include galactic fountains, as for the example dwarf galaxy at $z=0.2$ (see Figure~\ref{fig:MaNGAMaps}, middle panel, first row). Furthermore inflowing gas from mergers can also offset the rotational motion in projection, see e.g. the gas inflow from the lower left corner for the example $\mathcal{M^{*}}$ galaxy at $z=0.4$ (Figure~\ref{fig:MaNGAMaps}, right panel, third row). Similarly, cosmic gas inflows can also cause misalignment, as for the example massive dwarf galaxy at $z=0.0$ (Figure~\ref{fig:MaNGAMaps}, left panel, second row). Finally, the entire gas disc can be misaligned so that misalignment stems from the rotational motion rather than gas inflows or outflows, as in the example dwarf galaxy at $z=0.0$ (Figure~\ref{fig:MaNGAMaps}, left panel, first row). This shows that kinematically misaligned ionized gas is not necessarily due to feedback activity alone, but misalignment still correlates with enhanced feedback (since inflows and mergers can also trigger feedback events).

To assess the relationship between overmassive BHs and kinematic misalignment more quantitatively, we systematically compare kinematic offsets for the undermassive ($M_\mathrm{BH}/M_\mathrm{gas}$ below $\mathrm{P}_{20}$) and the overmassive sample ($M_\mathrm{BH}/M_\mathrm{gas}$ above $\mathrm{P}_{80}$) in Figure~\ref{fig:PADistribution}. We plot the distributions of $\Delta \mathrm{PA}$ for the edge-on, $\theta=0\degree$ case (solid lines) and the $\theta = 45 \degree$ case (dashed lines), with the shaded grey area indicating the misaligned regime. From left to right, the panels show the three different redshifts and from top to bottom, the panels show the three different stellar mass bins (as indicated by the plot titles). The plot titles give the sizes of the samples of galaxies that are sufficiently resolved and sufficiently rotationally supported for projections at $\theta=0\degree$ and $\theta = 45 \degree$, respectively.

We note that the massive dwarfs and $\mathcal{M^{*}}$ galaxies sample sizes increase with decreasing redshift, whilst the size of the dwarf galaxies sample decreases. This is due to the above mentioned resolution criterion (at least 50 star particles and 50 ionized gas cells within $1.5r_\mathrm{eff}$) which forces us to discard an increasing number of dwarf galaxies as the amount of ionized gas decreases with redshift. Comparing the sample sizes for the two inclination angles, we can see that the $\theta = 45 \degree$ sample size is greater than or equal to the $\theta = 0 \degree$ sample size in all cases. Again this is caused by the above mentioned resolution issues. When the galaxy is inclined towards the observer, we are likely to have a higher amount of gas in the l.o.s. due to the presence of outflows, so a higher number of galaxies will fulfill the resolution criterion at higher inclination.

Focusing first on the $\theta = 0 \degree$ projections, we can see that in all cases, the galaxies above $\mathrm{P}_{80}$ are either more likely (or for the $\mathcal{M^{*}}$ galaxies at $z=0.4$ equally likely, though note the small sample size) to be categorised as kinematically misaligned. Furthermore, these galaxies are also more likely to have extreme misalignments ($60 \degree \leq \Delta \mathrm{PA} < 120 \degree$).

Moving on to the $\theta = 45 \degree$ projections, we first note that in all cases these are more likely to be misaligned (or for the $P_{20}$ $\mathcal{M^{*}}$ galaxies at $z=0.4$ equally likely) compared to their $\theta = 0 \degree$ counterparts, as it is easier for inflows or outflows to offset the projected gas kinematics when the galaxy is inclined towards the observer.

Comparing the $P_{20}$ sample to the $P_{80}$ sample at $\theta = 45 \degree$ yields a more complex picture. In the majority of cases (six out of nine), the overmassive BHs are still more likely to be misaligned. However, in the remaining three cases, the $P_{20}$ galaxies are more or equally likely to be misaligned. 

We interpret these distributions by noting that overmassive BHs are associated with both extreme mass inflow and outflow rates (see Figure~\ref{fig:PhysicalPropsBH}) as well as higher outflow velocities. The higher outflow velocities make it easier to offset the rotational motion in projection leading to significant kinematic offsets. As we incline the galaxy disc towards the observer, the l.o.s. component of the rotational velocity decreases so outflows moving at slower velocities can also offset the rotational motion in projection, and the distinction between undermassive and overmassive BHs becomes less clear. We tested this hypothesis by inspecting the distribution of kinematic offsets split by $v_\mathrm{out}$ instead of $M_\mathrm{BH}/M_\mathrm{gas}$ and found that we obtain a similar distribution, demonstrating that outflow velocity is an important factor for shaping the $\Delta \mathrm{PA}$ distribution - though outflows are not the only process generating kinematically misaligned gas (see Figure~\ref{fig:MaNGAMaps}).

Furthermore, observed differences between the distributions of $\Delta \mathrm{PA}$ for AGN and non-AGN galaxies are also influenced by the inclination angle. We estimated inclination angles of the MaNGA galaxies from \citet{Penny2018a} using the $b/a$ axial ratios at $r_\mathrm{eff}$ from the NASA-Sloan Atlas catalogue. We found that the median inclination angle (measured with respect the edge-on configuration) for the dwarfs without AGN signatures is $\theta \sim 46 \degree$, whilst the median inclination angle for the dwarfs with AGN signatures is $\theta \sim 55 \degree$, so this might slightly increase the difference in kinematic offsets between non-AGN and AGN dwarfs (though note that this difference in inclination angles is much smaller than the difference between the two angles we consider here). Also note that \citet{Penny2018a} only considered quiescent dwarf galaxies. We do not impose a star formation based cut on our sample as this would leave us with too few objects, so this is a potential disparity as kinematic misalignment due to stellar feedback is likely more prominent in our sample.

The observed higher incidence of kinematically misaligned ionized gas for dwarf galaxies with AGN signatures could be caused by AGN-boosted outflow velocities \citep[see also][]{Duckworth2020b}, but also by other physical processes like cosmic inflows or mergers. Moreover, spectroscopic AGN signatures are easier to identify for galaxies which are inclined towards the observer, further increasing the likelihood of kinematic misalignment. To sum up, we find that kinematically misaligned gas is correlated with overmassive BHs but not always caused by AGN feedback itself.

\subsection{Radiative properties: mock X-ray luminosities}\label{subsec:xrayprops}
\begin{figure*}
    \centering
    \includegraphics[width=\textwidth]{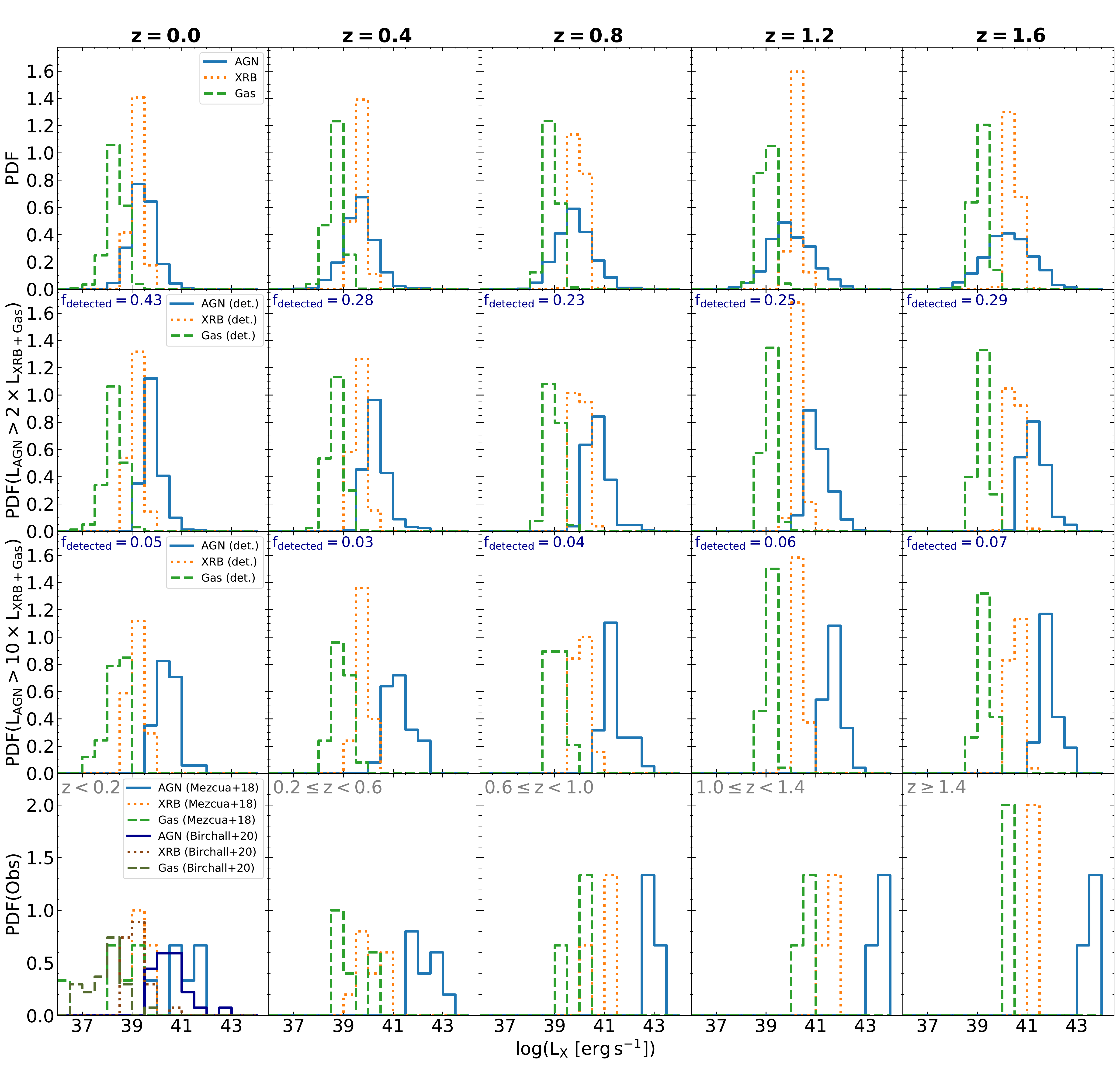}
    \caption{Distribution of the X-ray luminosities of BHs, XRBs, and hot gas across different redshifts. The top row shows the X-ray luminosity distributions for all \fable \ dwarf galaxies ($9.0 \leq \log(M_\mathrm{stellar} \ [\Msun]) < 9.5$). 
    The second row only includes the \fable \ dwarfs that fulfil the AGN selection criterion from \citet{Birchall2020}: $L_\mathrm{AGN}> 2 \times L_\mathrm{XRB+Gas}$. We also examine the X-ray luminosity distributions using an even stricter criterion ($L_\mathrm{AGN}> 10 \times L_\mathrm{XRB+Gas}$) in the third row. In both cases, we give the fraction of \fable \ dwarf galaxies with BHs that would get categorised as AGN using these criteria in the upper left hand corner. With the \citet{Birchall2020} criterion, a high fraction of BHs in dwarf galaxies are categorised as AGN, demonstrating that contamination from XRBs and hot gas should likely not pose a significant issue, even at the low-mass end. For comparison, we also show the X-ray distributions of the observed dwarf galaxies with AGN from \citet{Mezcua2018} and \citet{Birchall2020} in the bottom row. The local detections are in broad agreement with the \fable \ distributions, though the observed AGN are shifted towards higher luminosities. From the \fable \ luminosity distributions, we would expect many more AGN in dwarf galaxies to be detected by future X-ray surveys with higher sensitivities.}
    \label{fig:X-rayDistributions}
\end{figure*}

\begin{figure}
    \centering
    \includegraphics[width=\columnwidth]{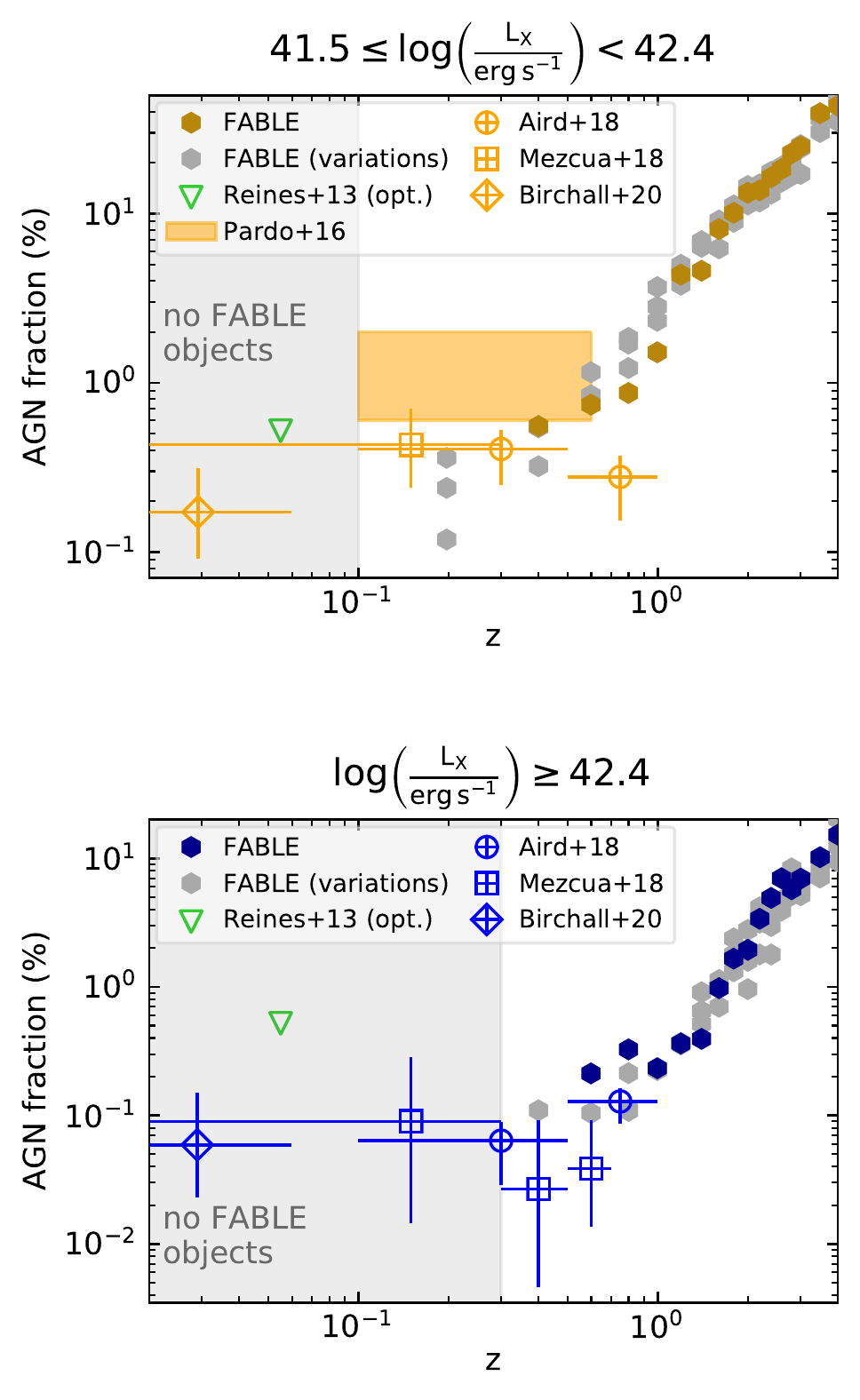}
    \caption{AGN fraction in dwarf galaxies ($9.0 \leq \log(M_\mathrm{stellar} \ [\Msun]) < 9.5$) as a function of redshift for two X-ray luminosity bins: $3.7 \times 10^{41} \ \mathrm{erg \, s^{-1}} \leq L_\mathrm{X} < 2.4 \times 10^{42} \ \mathrm{erg \, s^{-1}}$ (upper panel) and $L_\mathrm{X} \geq 2.4 \times 10^{42} \ \mathrm{erg \, s^{-1}}$ (lower panel). The AGN fractions for the fiducial \fable \ run are shown as dark orange and dark blue hexagons, respectively. Furthermore, we plot the AGN fractions of three of the \fable \ calibration runs, which tested different variants of the AGN feedback model, as grey hexagons. Observational data are also shown for comparison, as indicated by the legend. At low redshifts there are no \fable \ objects in either of the X-ray luminosity bins, suggesting that the \fable \ feedback model might suppress the occurrence of these highly-accreting objects in the local Universe. At intermediate redshifts,  \fable \ is in agreement with observed X-ray AGN fractions. For $z>1$, the simulation predicts that the AGN fraction should rapidly increase in both of the luminosity bins. Figure adapted from \citet{Mezcua2018} and \citet{Birchall2020}.}
    \label{fig:AGNFraction}
\end{figure}

As discussed in Section~\ref{sec:intro}, there have been a number of systematic X-ray searches for AGN in dwarf galaxies. \citet{Mezcua2018} presented a sample of 40 AGN candidates in dwarfs at $z \lesssim 2.4$ drawn from the \textit{Chandra} COSMOS Legacy Survey. Recently, \citet{Birchall2020} found 61 AGN candidates in local dwarf galaxies ($z \lesssim 0.25$) using the common footprint of the MPA-JHU catalogue and XMM DR7. Here we compare the X-ray properties of the \fable \ dwarf galaxies to the these observations, focusing on the redshift evolution and the detectability of these BHs in dwarf galaxies using X-ray surveys.

Current X-ray studies searching for AGN in dwarf galaxies face two main challenges. Since BHs in dwarf galaxies have relatively low masses ($M_\mathrm{BH} \sim 10^{5}\ \Msun$), their luminosities are also correspondingly lower. This makes it difficult to identify AGN in dwarf galaxies, unless they have a high Eddington fraction. For example, \citet{Mezcua2018} estimate that 95 per cent of their AGN candidates have near to super Eddington accretion rates ($f_\mathrm{Edd}>10^{-2}$). Contrast this with Section~\ref{subsec:population}, where we found that only one to ten per cent of \fable \ low-mass galaxies have BHs with $f_\mathrm{Edd}>10^{-2}$ between $z=0$ and $z=1$, though due to the efficient SN feedback at the low-mass end in \fable, this might be a lower limit.

\citet{Birchall2020}, on the other hand, have a higher proportion of low-luminosity AGN in their sample, with a typical Eddington fraction of $f_\mathrm{Edd} \sim 10^{-3}$, as they focus on local galaxies. Furthermore, their technique for separating AGN from other X-ray sources might allow them to identify more AGN sources (see below).

The second main issue for X-ray surveys is contamination from XRBs as well as hot gas emission. Due to the lower AGN luminosities, there is an overlap between the luminosity distributions of the contaminants and the AGN. In observational studies, it is therefore necessary to estimate the contributions from XRBs and hot gas to check that the integrated luminosity is significantly higher than the expected contribution from non-AGN sources.

\citet{Birchall2020} use the relation from \citet{Lehmer2016} to estimate the XRB contribution based on $M_\mathrm{stellar}$ and the SFR. For the hot gas contribution, they use the relation from \citet{Mineo2012} which gives the X-ray luminosity of the gas based on the SFR. They require that the total X-ray luminosity has to be three times higher than the estimated luminosity of the non-AGN sources (i.e. the putative AGN luminosity, $L_\mathrm{AGN}$, has to be twice as high as the luminosity of the contaminants, $L_\mathrm{XRB+Gas}$). This leaves them with 61 AGN candidates out of the 86 X-ray active dwarf galaxies in their original sample. Note that in their sample, galaxies with significantly higher X-ray luminosities than expected from XRBs also exceed the expected contribution from hot gas, so the latter is less of an issue when identifying AGN.

The AGN candidates from \citet{Mezcua2018} are drawn from a parent sample of $\sim 2300$ X-ray-selected type-2 AGN out to $z \sim 3$ \citep[see][]{Suh2017}. These source were classified by their spectroscopic type when available, or the photometric type. \citet{Mezcua2018} also check the contamination from XRBs using the \citet{Lehmer2016} relation. They find that the observed X-ray luminosities are more than $\sim 6 \sigma$ larger than expected from XRBs, and using the \citet{Mineo2012} relation, they find that the X-ray luminosities of all their sources are more than $\sim 34 \sigma$ above the luminosity expected from hot gas. So there is an extremely low chance of contamination - which is to be expected given the high Eddington fractions of the AGN candidates in this sample.

To study the redshift evolution of AGN and contaminant luminosities in \fable, we plot the PDFs of the AGN, XRB, and hot gas X-ray luminosities in Figure~\ref{fig:X-rayDistributions}. We use the bolometric corrections from \citet{Shen2020} to estimate the \fable \ X-ray AGN luminosities in the $0.5 - 10 \ \mathrm{keV}$ band, matching the band used by \citet{Mezcua2018}. In analogy to the observational studies, we estimate the XRB and hot gas contributions from the relations by \citet{Lehmer2016} and \citet{Mineo2012}, respectively. See Section~\ref{subsec:Xraymethods} for more details on the calculation of the different X-ray luminosities.

The top row of Figure~\ref{fig:X-rayDistributions} shows the luminosity distributions of all \fable \ dwarfs. There is a significant overlap between the luminosity distributions of AGN and XRBs. For the local galaxies, the contamination from XRBs is less significant as the sSFR is significantly lower (see Figure~\ref{fig:PhysicalPropsBH}). Note that for all redshifts, the hot gas and the AGN contribution are well separated, with the hot gas distribution shifted towards significantly lower luminosities than the XRB contribution. In agreement with the observational studies, we therefore conclude that the hot gas contribution should not pose a significant issue for the identification of AGN in dwarf galaxies.

In the second row, we show the distribution of the X-ray luminosities for all \fable \ dwarf galaxies that fulfil the \citet{Birchall2020} criterion ($L_\mathrm{AGN}> 2 \times L_\mathrm{XRB+Gas}$). In the top left corner, we give the fraction of dwarf galaxies with BHs, $f_\mathrm{detected}$, that fulfil this criterion. As expected from the full distribution, the highest fraction of AGN in dwarf galaxies are recovered for $z=0$ with $f_\mathrm{detected}=0.43$.  This then decreases with redshift to $f_\mathrm{detected}=0.23$ at $z=0.8$ and then increases again, as the mean AGN luminosities move towards higher values with higher redshift, to $f_\mathrm{detected}=0.29$ at $z=1.6$. Note that since \citet{Birchall2020} identified 61 out of 86 X-ray active dwarf galaxies as AGN, they obtained an even higher AGN fraction using this criterion. This difference is likely linked to the harder X-ray band used in their study ($2 - 12 \ \mathrm{keV}$) as well as completeness limits. Note that their original sample is shifted towards higher X-ray luminosities compared to \fable \ and that it only has a few dwarf galaxies with X-ray luminosities below $10^{39} \ \mathrm{erg \, s^{-1}}$.

In the third row, we show the distribution of the \fable \ galaxies with an even stricter criterion of $L_\mathrm{AGN}> 10 \times L_\mathrm{XRB+Gas}$. Whilst this efficiently separates the distributions of the AGN and the contaminants, it only classifies between three and seven per cent of sources as AGN.

In the bottom row, we plot the observed distributions from \citet{Mezcua2018} and \citet{Birchall2020}. We show the \citet{Mezcua2018} galaxies in five different redshift bins, approximately corresponding to the \fable \ redshifts considered. The last bin also includes a few higher redshift objects, as there is only a small number of AGN candidates at $z\geq 1.4$. The \citet{Birchall2020} galaxies are plotted in the $z<0.2$ panel, as all of their sources are at low redshifts, again note that the X-ray luminosities from \citet{Birchall2020} are based on a harder X-ray band ($2 - 12 \ \mathrm{keV}$), so these distributions are not directly comparable.

We only include observed dwarf galaxies with $M_\mathrm{stellar} \geq 10^{9} \ \Msun$ in this plot to match the \fable \ dwarf sample. The \citet{Mezcua2018} stellar masses were calculated from galaxy model templates based on the Chabrier IMF so we do not adjust these stellar masses. The stellar masses from \citet{Birchall2020} are derived using the Kroupa IMF, so we subtract 0.05 dex to adjust these to a Chabrier IMF.

The \citet{Mezcua2018} AGN luminosities are clearly separated from the contaminant distributions and more akin to the \fable \ distributions using the stricter separation criterion of $L_\mathrm{AGN}> 10 \times L_\mathrm{XRB+Gas}$. Though note that towards higher redshifts ($z\gtrsim1.0$) the observed AGN luminosities have very high values ($L_\mathrm{AGN} \sim 10^{43} \ \mathrm{erg \, s^{-1}}$), demonstrating that the observations mainly pick up a few very bright objects. Interestingly, we have no \fable \ objects with $L_\mathrm{AGN} \gtrsim 10^{43} \ \mathrm{erg \, s^{-1}}$ at any of the redshifts considered here, indicating again that the observed dwarfs may be accreting more efficiently than in our sub-grid model. From our simulated X-ray luminosity distributions, we would expect many more candidates to be identified by future X-ray surveys with lower flux limits.

The local \citet{Birchall2020} AGN distribution is less clearly separated from the contaminants than the \citet{Mezcua2018} dwarfs. Though the AGN luminosity distribution is still shifted towards higher luminosities than the equivalent \fable \ distribution. We also note that for both observational studies, there are no AGN sources with luminosities below $10^{39} \ \mathrm{erg \, s^{-1}}$. This again suggests that AGN accreting at low Eddington fractions are missed by these surveys.

We therefore conclude that sensitivity limits will be the main issue for further X-ray searches for AGN in dwarf galaxies. Whilst emission from XRBs and hot gas is less of an issue and a high fraction of BHs in dwarf galaxies should be identified despite contamination from these sources (between $\sim 20$ and $\sim 40$ per cent).

Though note that the relations used here to estimate the XRB and hot gas contribution are derived from massive galaxies. As dwarf galaxies tend to have sub-solar metallicities they might have an enhanced high-mass XRB population \citep[see discussion in e.g.][]{Mezcua2018,Birchall2020}. \citet{Lehmer2019} observe this enhancement for four dwarf galaxies, however, more data is needed to reliably estimate the XRB population in low-mass galaxies.

As X-ray searches are relatively unaffected by contamination from other sources, they can be used to establish an AGN fraction in different X-ray luminosity bins above the completeness limit of the respective survey. Next we compare these observed X-ray AGN fractions with AGN fractions from \fable. In Figure \ref{fig:AGNFraction}, we show the AGN fraction against redshift in two different X-ray luminosity bins: $3.7 \times 10^{41} \ \mathrm{erg \, s^{-1}} \leq L_\mathrm{X} < 2.4 \times 10^{42} \ \mathrm{erg \, s^{-1}}$ (upper panel) and $L_\mathrm{X} \geq 2.4 \times 10^{42} \ \mathrm{erg \, s^{-1}}$ (lower panel).

For both of these luminosity bins, we plot the observed AGN fractions obtained by \citet{Aird2018a}, \citet{Mezcua2018}, and \citet{Birchall2020}. Though note again that the \citet{Birchall2020} luminosities were measured in a slightly different X-ray band (2 - 12 keV band) whilst the other two studies use the 0.5 - 10 keV band. As in the previous section, the \fable \ luminosities are based on the 0.5 - 10 keV band.

In addition, we show the AGN fraction obtained by \citet{Pardo2016a} in the upper panel. Though this AGN fraction should be taken as an upper limit due to the relatively small sample, with over half of the detected AGN having X-ray luminosities below $10^{41} \ \mathrm{erg \, s^{-1}}$. We also show the AGN fraction from \citet{Reines2013}. However, this fraction should also be taken as an upper limit as it was derived from optically-selected dwarf galaxies. We plot this fraction as a green down-pointing triangle for comparison in both panels.

We show the simulation data from the fiducial \fable \ run for the two luminosity bins as dark orange and dark blue hexagons, respectively. Here the AGN fraction is defined as the number of dwarfs with an overall X-ray luminosity $L_\mathrm{X}= L_\mathrm{AGN} + L_\mathrm{XRB+Gas}$ in the respective luminosity bin normalised by the total number of dwarfs at the given redshift. We include the luminosity of the contaminants for consistency with the observations, though for these high-luminosity sources this makes little difference as the AGN is by far the dominant contribution (see Figure~\ref{fig:X-rayDistributions}). Note that, as in the previous sections, we only consider central dwarf galaxies here so we might slightly overestimate the AGN fraction.

To get a sense of the uncertainty in the simulated AGN fractions, we also plot the AGN fractions obtained by three of the \fable \ calibrations runs, where multiple variations of the AGN feedback parameters were tested, including variants without a duty cycle and weaker radio feedback \citep[see Appendix A of][for details]{Henden2018a}. The AGN fractions from the runs with the alternative AGN feedback models are shown as grey hexagons. 

For the $3.7 \times 10^{41} \ \mathrm{erg \, s^{-1}} \leq L_\mathrm{X} < 2.4 \times 10^{42} \ \mathrm{erg \, s^{-1}}$ luminosity bin, there are no \fable \ dwarfs at $z=0.0$ for any of the AGN feedback variations. At $z=0.2$, all of the three variant AGN feedback models produce AGN in this bin, with AGN fractions from $\sim 0.1$ to 0.4 per cent. Although note that in absolute numbers, this corresponds to one to three AGN in this luminosity bin, so these AGN fractions will be affected by small-number statistics. The fiducial \fable \ run only has AGN in this luminosity bin from $z=0.4$ on-wards. We shade the plot area corresponding to $z\leq0.1$ grey to indicate that we have no \fable \ objects in this redshift range.

The absence of these objects could be due the \fable \ outputs not being frequent enough\footnote{The \fable \ simulation outputs are spaced in redshift by $\Delta z =0.2$ for $z\leq3$ and by $\Delta z =0.5$ for $3 < z \leq4$.} as accretion rates are highly variable and AGN might only spend a short time in the high accretion state. Furthermore, it could be that we do not find any bright objects as the \fable \ sub-grid models prevent the AGN from reaching these high luminosities.

For the redshifts where there are both simulated and observed data points, $0.2 \leq z \leq 1$, the AGN fractions from \fable \ are in broad agreement with the observed fractions. Both the observed and the simulated data points are consistent with no significant evolution of the AGN fraction until $z \sim 1$. From $z=1$ on-wards, where we have no observational constraints, the simulations predict a significant rise in the AGN fraction from $\sim 1.5$ per cent at $z=1$ to $\sim 43$ per cent at $z=4$.  

Next, we focus on the highest luminosity bin $L_\mathrm{X} \geq 2.4 \times 10^{42} \ \mathrm{erg \, s^{-1}}$ in the lower panel. Here there are no \fable \ objects below $z=0.4$. At $z=0.4$, only one of the calibration runs has any AGN in this luminosity bins, with an AGN fraction of $\sim 0.1$ per cent, whilst the fiducial run only has AGN in this bin from $z=0.6$. For $0.4 \leq z \leq 1.0$ where we have both observational and simulated data points, these are in good agreement and both consistent with no significant evolution in the AGN fraction until $z \sim 1$. Similar to the other luminosity bin, however, we see a significant increase in bright AGN at high redshifts, increasing from $\sim 0.2$ per cent at $z=1$ to $\sim 15$ per cent at $z=4$. We caution, however, that these AGN fractions are sensitive to the seeding model and the assumption that every DM halo above $5\times 10^{10} \ h^{-1} \, \Msun$ should host a BH (see Section~\ref{subsubsec:SeedingAndGrowth} for details on the \fable \ seeding model).

The prediction that the high-luminosity AGN fraction in dwarf galaxies will significantly increase with redshift can be tested with upcoming X-ray missions. We estimate the X-ray luminosity limits across the redshift range using the flux limits for \textit{Athena}\footnote{\url{https://www.cosmos.esa.int/documents/400752/507693/Athena_SciRd_iss1v5.pdf}}, and \textit{Lynx}\footnote{\url{https://wwwastro.msfc.nasa.gov/lynx/docs/LynxInterimReport.pdf}}. For a given redshift, we calculate the luminosity limit as $L_\mathrm{X} = 4 \pi d^{2}_\mathrm{L}f_\mathrm{X}$, where $d_\mathrm{L}$ is the luminosity distance and $f_\mathrm{X}$ is the flux limit. Both of the above instruments are more sensitive in the soft band (0.5 - 2 keV) than in the hard band (2 - 10 keV). Therefore, we use the soft and hard band flux limits to obtain lower and upper limits for detectable X-ray luminosities, respectively.

We find that \textit{Athena} should be able to measure the AGN fraction for $L_\mathrm{X} \geq 2.4 \times 10^{42} \ \mathrm{erg \, s^{-1}}$ up to $z\sim 1.5 - 3.2$. The AGN fraction for $3.7 \times 10^{41} \ \mathrm{erg \, s^{-1}} \leq L_\mathrm{X} < 2.4 \times 10^{42} \ \mathrm{erg \, s^{-1}}$ should be measurable up to $z \sim 0.7 - 1.5$. Therefore, \textit{Athena} should be able to detect the upturn in the AGN fraction at $z\gtrsim 1.0$ for the high-luminosity AGN, and it might also provide further constraints for the other luminosity bin.

\textit{Lynx}, however, will be able to constrain the AGN fraction across the whole redshift range ($0 \leq z \leq 4$) for both of the luminosity bins considered here. At $z=4$, we find that the soft X-ray luminosity limit is $L_\mathrm{X,soft} \sim 1.3 \times 10^{40} \ \mathrm{erg\,s^{-1}}$ and the hard X-ray luminosity limit is $L_\mathrm{X,hard} \sim 1.6 \times 10^{40} \ \mathrm{erg\,s^{-1}}$, so \textit{Lynx} will be able to observe the evolution of the AGN fraction even for low-luminosity AGN.

\section{Discussion} \label{sec:discussion}
\subsection{Feedback processes in dwarf galaxies}

Dwarf galaxies are an important testbed of galaxy formation and cosmology. Notwithstanding massive theoretical effort, it remains challenging to match the observed abundances and structural properties of dwarfs, requiring either to revisit our physical models or to even think outside of the standard $\Lambda$CDM framework. For example, the majority of large-scale galaxy formation simulations employ strong SN feedback to reproduce the low-mass end of the observed GSMF \citep[see e.g.][]{Bower2012,Dubois2014a,Vogelsberger2014, Schaye2015,Pillepich2018}. However, this approach is called into question by high-resolution (zoom-in) galaxy formation simulations (as well as small scale simulations of the ISM), which find it difficult to regulate star formation with SN feedback alone \citep[e.g.][]{Hopkins2014,Kim2015a,Kimm2015,Hu2016b,Hu2017,Marinacci2019,Smith2018b} so that other sources of star formation regulation need to be invoked such as stellar winds, photo-heating, radiation pressure effects, cosmic rays and (MHD) turbulence.

Interestingly, when some of these additional physical processes are implemented alongside SN feedback, stellar feedback still struggles to regulate star formation in dwarf simulations \citep[e.g.][]{Kimm2018}. To obtain efficient feedback in a cosmological setting, stellar feedback then needs to be enhanced, e.g. by boosting SN rates \citep{Rosdahl2018} or via high star formation efficiencies \citep{Hopkins2018a}. 

While strong stellar feedback seems required to match the observed dwarfs, it is likely to stunt AGN growth and feedback in these low-mass systems if a considerable gas reservoir is removed from the innermost regions \citep[see e.g.][]{Angles-Alcazar2017, Habouzit2017,Trebitsch2018,Habouzit2020}. Moreover, it is worth noting that most galaxy formation simulations use the Bondi-Hoyle-Lyttleton rate to model BH accretion, which suppresses BH growth close to the seed mass due to its quadratic dependency on $M_\mathrm{BH}$. In cosmological simulations, seed masses of $\sim 10^{5} - 10^{6} \ \Msun$ are typically employed, so that the slow growth phase then by construction of these models corresponds to the dwarf regime. 

\subsection{Comparison with other works}

The suppression of BH growth in dwarf galaxies (both by strong SN feedback and the Bondi-like accretion prescription) has meant that so far AGN feedback in dwarfs has largely been neglected by simulators. Novel observations of AGN in dwarf galaxies question these assumptions, as contrary to the common theoretical models, there is a population of dwarf galaxies where AGN can accrete efficiently. 

Recent theoretical works have started exploring BH activity in the dwarf regime using analytical tools \citep{Silk2017,Dashyan2018}, isolated simulations \citep{Koudmani2019}, cosmological zoom-in simulations \citep{Trebitsch2018, Barai2018b, Bellovary2019} as well as cosmological boxes \citep{DiCintio2017,Habouzit2017, Sharma2020}. 

\fable \ is a large-scale cosmological simulation with a $\sim 60 \ \mathrm{Mpc}$ box modelling a representative region of the Universe, including rare environments. This allows us to obtain a statistical sample of dwarf galaxies which can be compared to wide-field observational surveys: the \fable \ box contains $\lesssim 10^{3}$ central dwarfs (i.e. not satellites) for simulation outputs between $z=0$ and $z=1$ and due to the efficient seeding, at least 98 per cent of these central dwarfs host a black hole. The sample size and range of environments are the main strength of this approach, enabling us to reproduce the rare bright AGN in dwarfs. However, due to the \fable \ resolution, we cannot probe galaxies with stellar masses less than $M_\mathrm{stellar} = 10^{9} \ \Msun$, comparable in stellar mass to the Large Magellanic Cloud (LMC). Though note that most current observational investigations of AGN in dwarf galaxies are also (mainly) restricted to the LMC mass regime \citep{Reines2013,Penny2018a,Dickey2019a,Liu2020} as Eddington-limited AGN in more massive dwarfs are brighter and therefore easier to detect. Furthermore, AGN identification methods which are based on samples of massive galaxies (such as the BPT diagram) still apply in the LMC mass range, whilst this is no longer the case for lower-mass dwarf galaxies \citep[][also see Section~\ref{sec:intro}]{Cann2019}.

Other studies also analyse AGN feedback across a range of stellar masses down to the dwarf regime \citep[e.g.][]{Beckmann2017,Weinberger2018,Zinger2020}, however, the focus of these studies is mostly on the high-mass regime as special care needs to be taken to distinguish the AGN trends in the low-mass regime due to the strong supernova feedback employed by the majority of large-scale cosmological simulations. 

\citet{Sharma2020} present a similar study on AGN feedback in dwarfs based on the \textsc{Romulus25} simulation, using a 25 cMpc box compared to the $\sim 60 \ \mathrm{cMpc}$ box used for \fable, but with significantly higher resolution ($\gtrsim$ 100 higher DM mass resolution). These simulations also model the BH dynamics due to dynamical friction \citep{Tremmel2015}. Note that the BHs in dwarf galaxies in \textsc{Romulus25} are overmassive with respect to the observed scaling relations, whilst the BHs in the \fable \ dwarfs are undermassive compared to observations, so the results from \citet{Sharma2020} likely provide an upper limit for the impact of AGN feedback on dwarfs. Encouragingly, they find that dwarfs with increased BH growth have suppressed star formation rates, even at low redshifts. 

\subsection{Future prospects}

\subsubsection{Next-generation AGN models}

The next step is now to develop novel subgrid models for BH physics to study AGN activity much more realistically in a cosmological context. AGN feedback in dwarf galaxies is an exciting possibility as it is still under-explored theoretically, but very promising energetically \citep[see][]{Dashyan2018}.

Several BH seeding mechanisms should naturally predict BHs in dwarfs, for example remnants of Population III stars \citep[e.g.][]{Madau2001,Heger2003,Volonteri2003,Heger2010,Whalen2012,Karlsson2013} or dense nuclear star clusters \citep[e.g.][]{Begelman1978, PortegiesZwart2002,Freitag2005,Omukai2008,Devecchi2008,Katz2015}. Recent work has started exploring alternative seeding models in cosmological simulations, growing BHs from smaller seeds \citep[see e.g.][]{Habouzit2017,DeGraf2020}. However, observations cannot yet clearly distinguish between different seeding models \citep[e.g. see discussion in][]{Volonteri2008,Greene2019,Mezcua2019b} and the true occupation fraction of BHs in dwarf galaxies is still unknown (see Section~\ref{sec:intro}).

To investigate AGN in dwarf galaxies, in addition to improved seeding models, it will be necessary to explore accretion mechanisms beyond the widely adopted (yet very simplistic) Bondi model, with several models considered, such as a supply-limited accretion scheme \citep{Beckmann2018a,Beckmann2018b}, torque-driven BH growth \citep{Angles-Alcazar2015,Angles-Alcazar2017a} or accretion disc models \citep[e.g.][]{Power2010,Fiacconi2018,Bustamante2019}.

The modelling of the different AGN feedback channels is equally important. Here, with \fable, we only consider simple, isotropic feedback, however observations find evidence for bipolar outflows from AGN \citep[e.g.][]{Rupke2011,Maiolino2012}. Note that AGN feedback in \fable \ is modelled as mechanical feedback and a radiation field around the BH (see Section~\ref{subsubsec:AGNfeed}), but other processes that are not included here could also play an important role, such as outflows driven by radiation pressure \citep[e.g.][]{Bieri2017a,Costa2018,Costa2017} or jet-driven outflows. With a more realistic AGN feedback model, \fable \ underestimating the BH luminosity would have an impact beyond just underestimating the mechanical AGN feedback and photoionization, since increasing the BH luminosity would also e.g. increase radiation pressure.

Even though \fable \ under-predicts AGN activity in dwarf galaxies, as evidenced by the comparison with observed scaling relations and high-luminosity X-ray AGN fractions, there is still an effect by AGN feedback on dwarf galaxy properties in \fable. This suggests that in reality the impact of AGN feedback on dwarfs could be even more significant \citep[with the caveat that whilst the \fable \ BH growth is inefficient, the seeding is very efficient, see e.g.][]{DeGraf2020}. Taking these results at face value, it is then inevitable to conclude that the effects of AGN feedback on the evolutionary history of dwarfs need to be explored, which may lead to re-evaluation of stellar feedback models as well.

To aid the development of theoretical models, it will be crucial to assess how many dwarfs harbour central BHs and how these BHs grow through cosmic time. Future observational facilities like \textit{Lynx} or LISA will provide important observational constraints, as these instruments will probe the evolutionary history mapping both BH seeding and BH growth.

\subsubsection{The `cusp vs. core' problem}

AGN feedback in dwarf galaxies might also contribute to the resolution of an on-going debate about the so-called `cusp vs. core' problem. This controversy stems from the observed rotation curves of some dwarf galaxies, which seem to suggest cored DM density profiles rather than the cuspy profiles predicted by $\Lambda$CDM \citep{Flores1994,Moore1994,DeBlok2001,Gentile2004,Walker2011}. Currently there is no consensus about the origin of the diversity of inferred DM density profiles. Some suggest that the inferred cored profiles could be due to uncertainties in circular velocity measurements \citep[e.g.][]{Marasco2018,Oman2019}. Others have questioned the validity of $\Lambda$CDM and come up with alternative DM models - such as warm DM \citep[e.g.][]{Lovell2012} or self-interacting DM \citep[e.g.][]{Vogelsberger2014a}. Yet others have invoked baryonic processes transforming cusps to cores via strong feedback \citep[see e.g.][]{Navarro1996,Governato2010,DiCintio2013,Onorbe2015,Fitts2017}. 

In \fable, we find no significant differences in DM profiles for dwarf galaxies with overmassive BHs (not shown here), despite the clear impact on the baryonic component. This indicates that the \fable \ AGN feedback is not strong enough to influence the DM component. Furthermore, this is likely also compounded with the resolution of the simulations. However, given that \fable \ likely underestimates the luminosities of AGN in dwarf galaxies, AGN feedback may still play a role in transforming cusps into cores in dwarfs.

\section{Conclusions} \label{sec:conclusions}
Recent observations have uncovered a population of dwarf galaxies hosting AGN, providing tantalizing hints that AGN feedback could also play a role in low-mass galaxies (see Section~\ref{sec:intro}). We used the cosmological simulation suite \fable \ to investigate the impact of AGN feedback on (central) low-mass galaxies ($10^{9.0} \ \Msun \leq M_\mathrm{stellar} < 10^{10.5} \ \Msun$), with a particular focus on dwarf galaxies ($10^{9.0} \ \Msun \leq M_\mathrm{stellar} < 10^{9.5} \ \Msun$). We examined the distribution of bolometric BH luminosities (Section~\ref{subsec:population}) as well as the scaling relations (Section~\ref{subsec:scalings}). Furthermore, we studied the correlations between overmassive BHs and host galaxy properties, in particular outflows, in Section~\ref{subsec:offsetBHs}. We also compared the \fable \ low-mass galaxies to observational studies by creating mock MaNGA l.o.s. velocity maps (Section~\ref{subsec:MockMaNGAObs}). Moreover, we estimated the X-ray luminosities for AGN, XRBs and hot gas to assess the detectability of AGN in dwarf galaxies with wide-field X-ray surveys (Section~\ref{subsec:xrayprops}). Our most important findings are the following:

\begin{enumerate}
	\item The majority of AGN in low-mass galaxies, and in particular in dwarf galaxies, are outshone by the stellar component. This renders these AGN difficult to detect. However, the high-redshift regime is more promising as more low-mass galaxies reach high Eddington fractions and luminosities.
	\item The $M_\mathrm{BH} - M_\mathrm{stellar}$ scaling relation for \fable \ low-mass galaxies is broadly in agreement with observed late-type scaling relations, though note that the \fable \ relation is undermassive compared to most of the observed relations. This indicates that SN feedback may be too strong in the low-mass systems stunting BH growth.
	\item Low-mass galaxies with large positive offsets from the $M_\mathrm{BH} - M_\mathrm{gas}$ or the $M_\mathrm{BH} - M_\mathrm{stellar}$ relation have increased mass outflow and inflow rates. These overmassive BHs cause higher outflow velocities and hotter (albeit multiphase) outflows. While warm outflows typically generate galactic fountains, the hot outflow component is able to escape the host haloes, leading to a reduced gas reservoir. Future MUSE observations should be able to test our predicted outflow properties.
	\item We find that quenching of dwarfs proceeds via two channels. At lower redshifts ($z \lesssim 2$) where AGN accretion rates are low, SN feedback mainly regulates star formation in dwarf galaxies, whilst above $z \sim 2$, AGN feedback is strong enough to suppress star formation in dwarfs as well. We note that these conclusions are sensitive to the strong SN model adopted, with AGN quenching possibly more important at lower redshifts, too.
	\item Galaxies with overmassive BHs are more likely to have an ionized gas component that is kinematically misaligned from the stellar component in mock MaNGA l.o.s. velocity maps in agreement with observations. While fast AGN-boosted outflows are partly responsible for this misalignment, we caution that other factors such as cosmic inflows or mergers are also important.
	\item X-ray surveys are a promising tool for identifying AGN in dwarf galaxies. Even at the low-mass end, the AGN, XRB and hot gas X-ray luminosity distributions are sufficiently separated to identify a high fraction of AGN in dwarfs. Sensitivity remains the main issue, with the observed X-ray distributions shifted towards higher luminosities than \fable. We predict that future X-ray surveys should uncover many more dwarf galaxies with AGN with lower Eddington fractions and at higher redshifts.
	\item By comparing the occupation fraction of luminous AGN in dwarf galaxies in \fable \ to observations, we find that luminous AGN are missing from \fable \ at low redshifts, again indicating SN feedback-starved BH growth. Good agreement is, however, reached for intermediate redshifts. For high redshifts ($z>1$), where there are currently no observational constraints, we predict that the fraction of luminous AGN in dwarf galaxies should rise rapidly. 
\end{enumerate}

Like the majority of galaxy formation simulations, \fable \ has been designed such that SN feedback regulates the low-mass end of the galaxy mass function. Notwithstanding this, we find that AGN feedback in \fable \ has a clear impact on the outflow properties of dwarf galaxies across cosmic time and contributes to quenching at high redshifts. Ongoing and upcoming state-of-the-art observations with e.g. JWST, MUSE, \textit{Athena}, \textit{Lynx}, ngVLA, and LISA will be able to probe the elusive AGN population in dwarfs to much lower luminosities and at higher redshifts. It is hence of paramount importance to develop next generation galaxy formation models with more realistic seeding models and more efficient AGN accretion in low-mass objects to explore this currently unknown territory, with the ultimate goal of elucidating the role of AGN in dwarf galaxies.

\section*{Acknowledgements}
We would like to thank Martin Haehnelt, Ewald Puchwein and Sylvain Veilleux for helpful comments. An anonymous referee provided a comprehensive
review for which we are grateful. SK, NAH and DS acknowledge support by the STFC and the ERC Starting Grant 638707 `Black holes and their host galaxies: co-evolution across cosmic time'. This work made use of the following facilities: the Cambridge Service for Data Driven Discovery (CSD3) operated by the University of Cambridge Research Computing Service (www.csd3.cam.ac.uk), provided by Dell EMC and Intel using Tier-2 funding from the Engineering and Physical Sciences Research Council (capital grant EP/P020259/1), and DiRAC funding from the Science and Technology Facilities Council (www.dirac.ac.uk), as well as the DiRAC@Durham facility managed by the Institute for Computational Cosmology on behalf of the STFC DiRAC HPC Facility (www.dirac.ac.uk). The equipment was funded by BEIS capital funding via STFC capital grants ST/P002293/1, ST/R002371/1 and ST/S002502/1, Durham University and STFC operations grant ST/R000832/1. DiRAC is part of the National e-Infrastructure.

\section*{Data availability}
The data underlying this article will be shared on request to the corresponding author.




\bibliographystyle{mnras}
\bibliography{Paper2} 




\appendix

\section{Overmassive black holes at fixed stellar mass} \label{appsec:offsetbhsstellar}
\begin{figure*}
	\includegraphics[width=\textwidth]{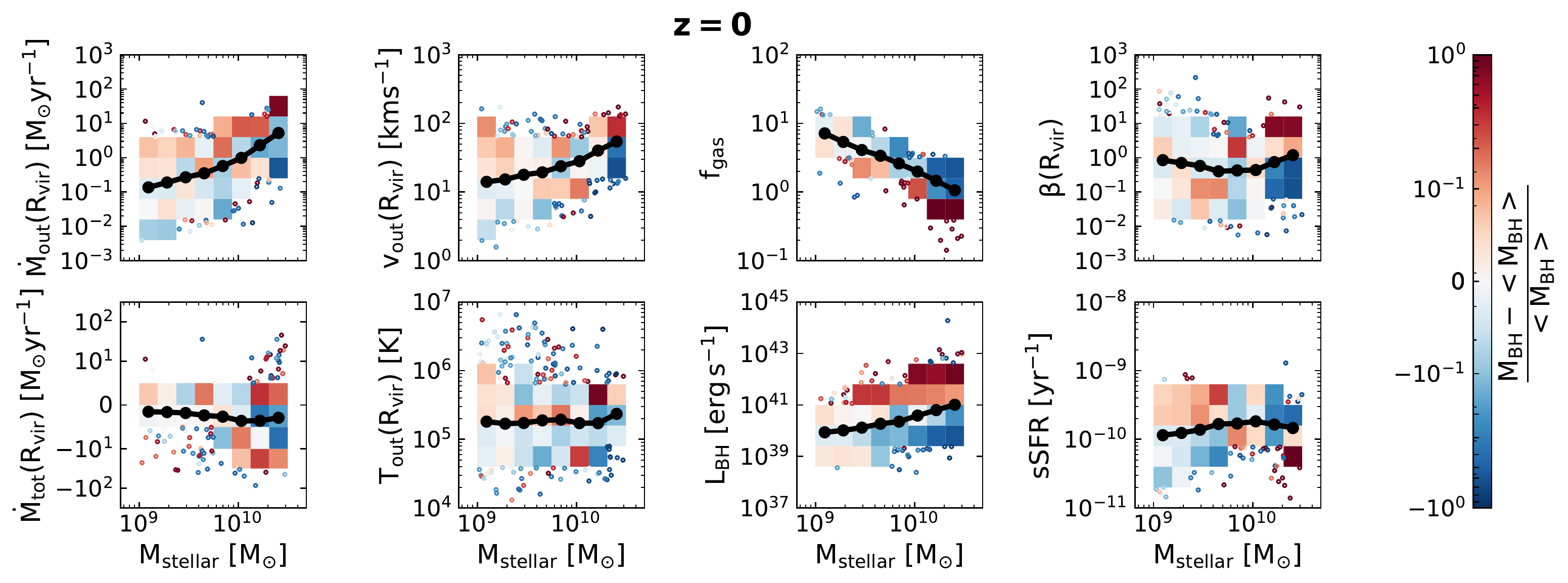}\\
	\includegraphics[width=\textwidth]{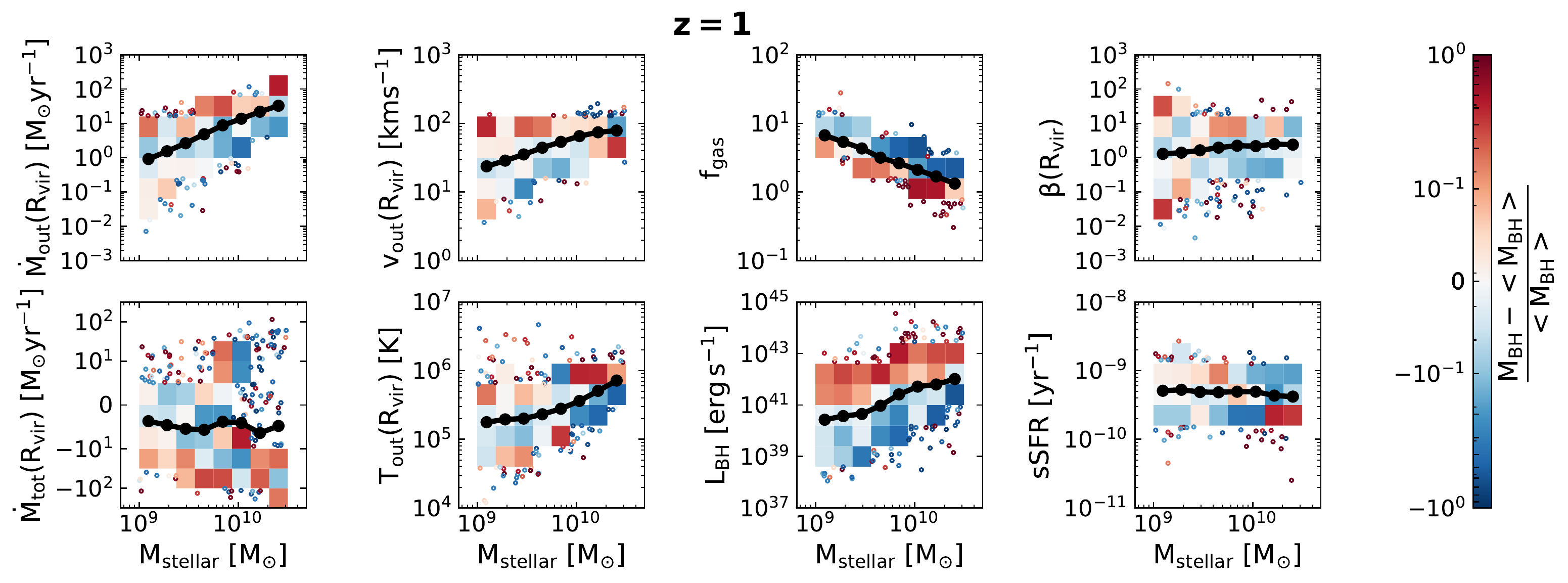}
    \caption{Outflow characteristics and galaxy properties against total stellar mass at $z=0$ (upper panel) and $z=1$ (lower panel) for the \fable \ low-mass galaxy sample. In each case the binned mean relation is shown as a solid black curve with the filled black circles indicating the bin midpoints. The colour coding of the distribution indicates the offset from the $M_\mathrm{BH} - M_\mathrm{stellar}$ scaling relation from Figure \ref{fig:ScalingRelations}, with blue for undermassive and red for overmassive BHs. 2D bins with at least ten objects are colour-coded according to the mean BH mass offset, and otherwise we plot the individual objects colour-coded by their respective BH mass offsets. Galaxies with overmassive BHs have increased outflow and inflow rates, leading to a bimodal distribution for the total mass flow rate. The outflows are faster and hotter in the overmassive regime, similar to the fixed gas mass case. Overmassive BHs at fixed $M_\mathrm{stellar}$ are also associated with reduced gas fractions and increased BH luminosities. The trends for the star-formation related properties (mass loading factor and sSFR), however are either washed out or inverted below $M_\mathrm{stellar} \lesssim 6 \times 10^{9} \ \Msun$ compared to the fixed gas mass plots. See Figure \ref{fig:PhysicalPropsBH} for corresponding plots at fixed gas mass.}
    \label{fig:PhysicalPropsBHStellar}
\end{figure*}

\begin{figure*}
    \centering
    \includegraphics[width=\columnwidth]{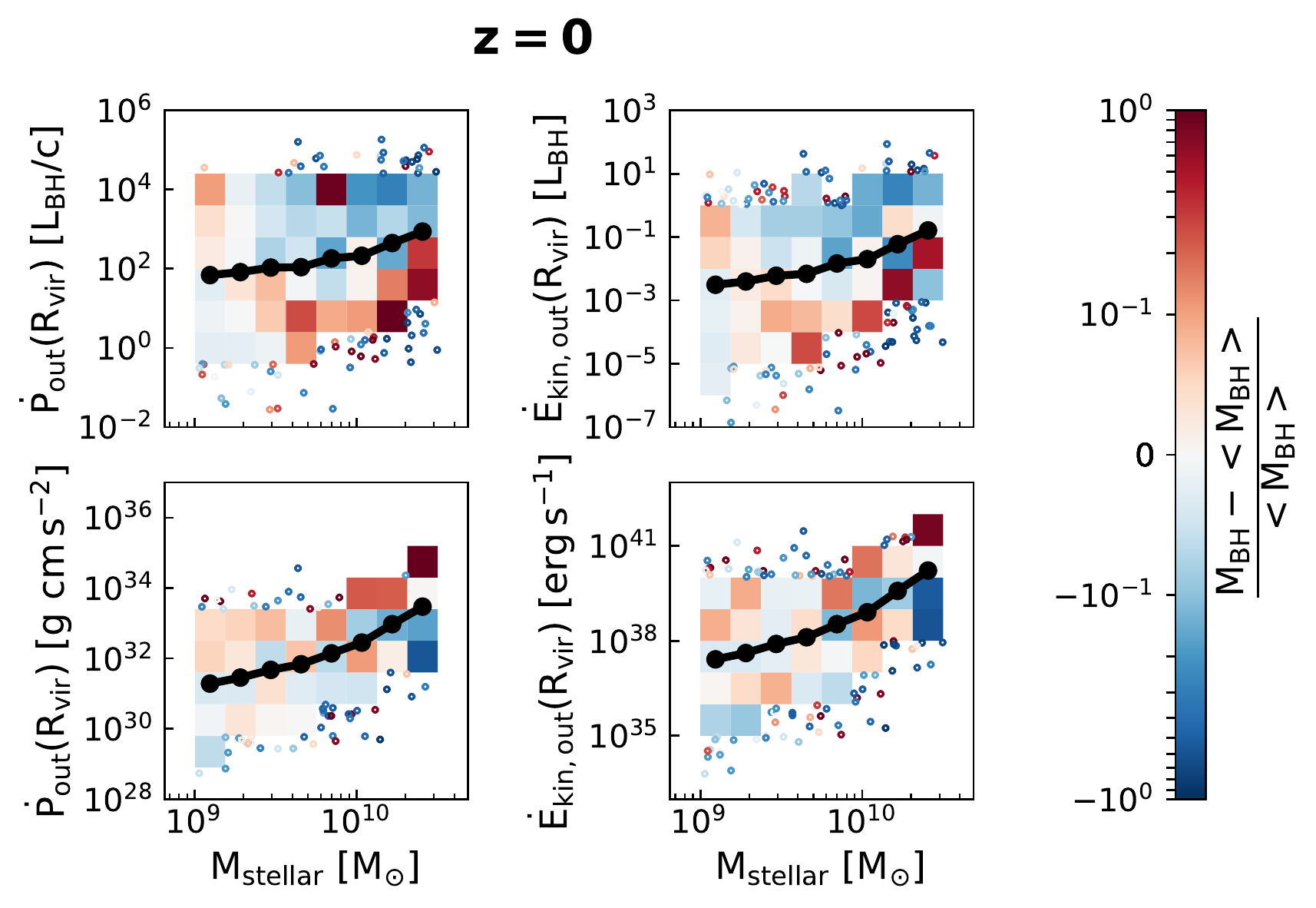}
    \includegraphics[width=\columnwidth]{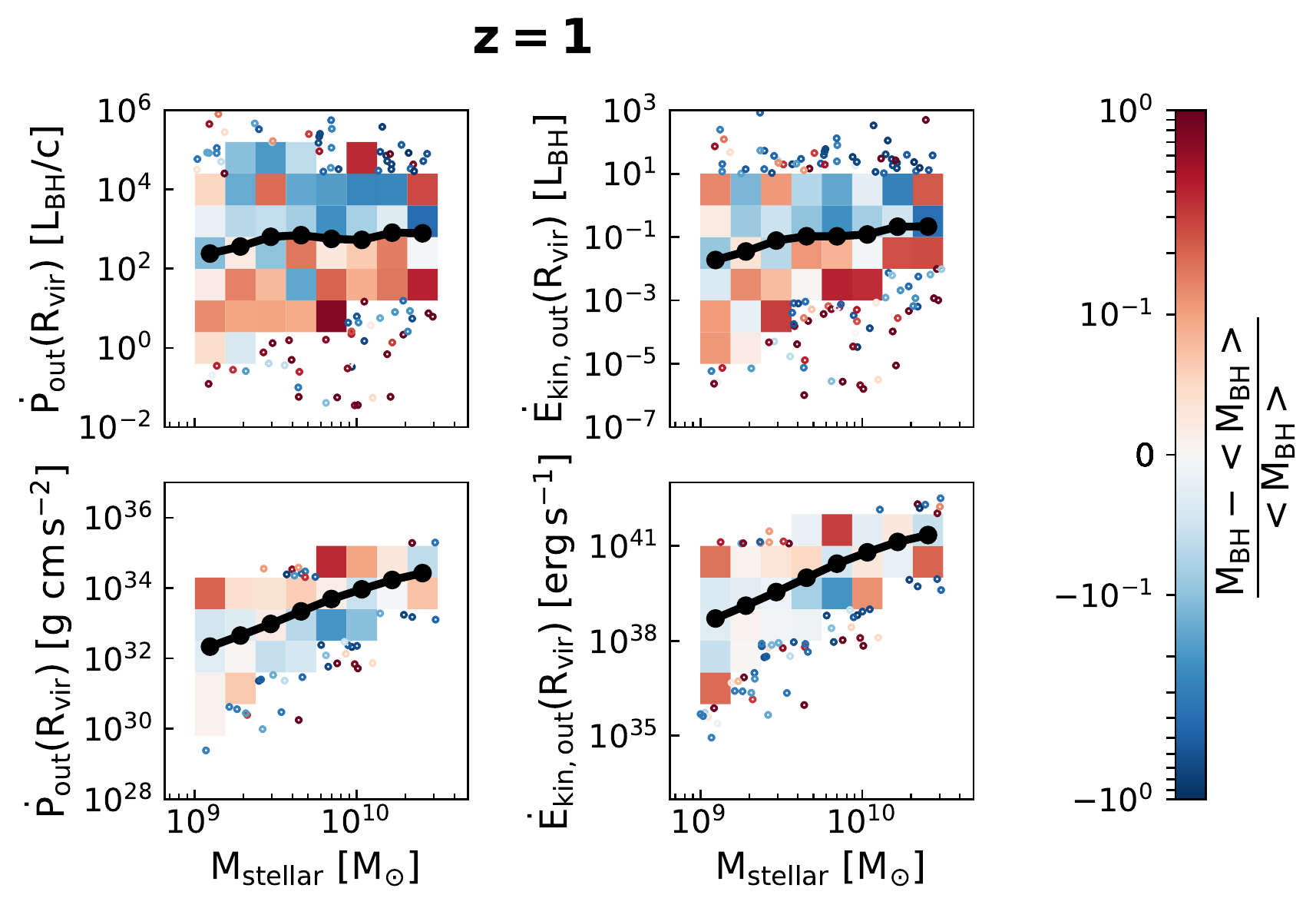}
    \caption{Momentum and kinetic energy outflow rates against total stellar mass at $z=0$ (left panel) and $z=1$ (right panel) for the \fable \ low-mass galaxy sample. The top row is normalised by the BH luminosity whilst the bottom row shows the outflow quantities in cgs units. The binned mean relations are shown as solid black curves, with the bin midpoints indicated by the filled black circles. Colour coding indicates the offset from the $M_\mathrm{BH} - M_\mathrm{stellar}$ scaling relation from Figure \ref{fig:ScalingRelations}, with blue for undermassive and red for overmassive BHs. Where there are fewer than ten objects per 2D bin, the individual objects are plotted. In cgs units, both momentum and energy outflow rates are clearly enhanced for overmassive BHs. However normalizing by the BH luminosity washes out this correlation at both redshifts due to strong coupling between BH masses and luminosities (see Figure \ref{fig:PhysicalPropsBHStellar}). See Figure \ref{fig:mom_and_egy_rates} for corresponding plots at fixed gas mass.}
    \label{fig:mom_and_egy_rate_stellar}
\end{figure*}

In Section \ref{subsec:offsetBHs}, we illustrated how the offset from the $M_\mathrm{BH} - M_\mathrm{gas}$ relation is correlated with the properties of the host galaxy, in particular with the outflow properties. We found that low-mass galaxies with overmassive BHs are associated with more powerful outflows, lower sSFRs and gas fractions, as well as higher BH luminosities. 

We focused on the effect of varying BH mass at fixed gas mass, as gas abundance is the key driver for BH activity. On the one hand gas provides the fuel for the BH to grow through accretion, and on the other hand a certain amount of gas is needed to be able to drive outflows. 

However note that overmassive BHs at fixed gas mass tend to also have larger stellar masses than their undermassive counterparts due to the tight relation between $M_\mathrm{BH}$ and $M_\mathrm{stellar}$ (see Figure \ref{fig:ScalingRelations}). To ensure that the increased outflow activity is not just driven by stellar feedback, we present plots analogous to the ones in Section \ref{subsec:offsetBHs}, but at fixed stellar mass. 

Figure \ref{fig:PhysicalPropsBHStellar}, in analogy to Figure \ref{fig:PhysicalPropsBH}, shows outflow quantities at $R_\mathrm{vir}$ (mass outflow rate $\dot{M}_\mathrm{out}$, total mass flow rate $\dot{M}_\mathrm{tot}$, outflow velocity $v_\mathrm{out}$, outflow temperature $T_\mathrm{out}$, mass loading factor $\beta$) as well as integrated galaxy properties (gas mass to stellar mass ratio $f_\mathrm{gas}$, BH luminosity $L_\mathrm{BH}$, sSFR) at $z=0$ (top panel) and at $z=1$ (bottom panel) plotted against stellar mass.

We plot the distribution of the whole low-mass galaxy population. The colour-coding indicates the offset from the $M_\mathrm{BH} - M_\mathrm{stellar}$ relation from Figure \ref{fig:ScalingRelations}, with red for overmassive and blue for undermassive BHs. Where there are at least ten objects per 2D bin we show the mean BH mass offset of the respective bin, and otherwise we plot the individual objects. Note that the scatter in BH masses at fixed stellar mass is much smaller than at fixed gas mass (see Figure \ref{fig:ScalingRelations}). We also plot the mean relation as a solid black curve, with the bin midpoints marked by filled black circles.

At fixed stellar mass, we recover the same outflow trends as at fixed gas mass, albeit with a weaker correlation due to the smaller scatter and differing gas reservoirs at fixed stellar mass. We find that overmassive BHs are correlated with both increased mass outflow rates and mass inflow rates, leading to a bimodal distribution for the total mass outflow rate (see Figure \ref{fig:PhysicalPropsBHStellar}, first column). 

Moving on to the second column, we find that overmassive BHs at fixed stellar mass tend to be associated with faster and hotter outflows, but the correlation is weaker than at fixed gas mass. We also inspected the outflow properties at smaller scales and found that there is a much stronger correlation between overmassive BHs and increased outflow velocities and temperatures at $0.2 R_\mathrm{vir}$. Similarly, we found that at fixed gas mass, the strength of the correlation between overmassive BHs and $v_\mathrm{out}$ and $T_\mathrm{out}$ is stronger at $0.2 R_\mathrm{vir}$. This suggests that the hot phase driven by the AGN is more dominant at smaller radii, indicating that some of the hot gas falls back as a galactic fountain.

The third column of Figure \ref{fig:PhysicalPropsBHStellar} shows that there is a clear correlation between overmassive BHs and suppressed gas fractions, in agreement with the results at fixed gas mass. Furthermore, overmassive BHs are generally associated with higher BH luminosities, though this trend is less significant for dwarf galaxies at $z=0$.

The last column shows galaxy properties related to star formation: the mass loading factor $\beta$ and the sSFR. Here we find the most significant difference between the trends at fixed gas mass and fixed stellar mass. Whilst at fixed gas mass, we found that overmassive BHs are associated with increased $\beta$ and suppressed sSFRs for the whole galaxy gas mass range, the trends are less clear at fixed stellar mass. Below $M_\mathrm{stellar} \lesssim 6 \times 10^{9} \ \Msun$, the correlation is either washed out or inverted. This suggests that for dwarf galaxies in \fable \ there is no strong connection between AGN activity and quenching at low redshifts. Though note that the sub-grid model might underestimate AGN luminosities in this mass range (see Figure \ref{fig:AGNFraction} in Section \ref{subsec:xrayprops}).

Finally, we also consider the momentum and kinetic energy outflow rates at fixed stellar mass in Figure \ref{fig:mom_and_egy_rate_stellar} (see Figure \ref{fig:mom_and_egy_rates} for the equivalent figure at fixed gas mass). We show these outflow rates at $z=0$ (left panel) and $z=1$ (right panel). The top row shows the momentum and energy outflow rates in units of the BH luminosity, whilst the bottom row shows the same quantities in cgs units. Again, the colour-coding indicates the offset from the $M_\mathrm{BH}$ - $M_\mathrm{stellar}$ relation. We show the mean bin values where there are at least ten objects per 2D bin, and otherwise plot the individual objects. The mean outflow rates are shown as solid black curves, with the bin midpoints marked by filled black circles.

Focusing first on the bottom row, we find that galaxies with overmassive BHs have increased momentum and energy outflow rates, at both redshifts. The correlation is less strong than at fixed gas mass - which we attribute again to the range of gas reservoirs at fixed $M_\mathrm{stellar}$ and the small scatter in the  $M_\mathrm{BH}$ - $M_\mathrm{stellar}$ relation. 

However, when we normalise the momentum and energy outflow rates by the BH luminosity, the relationship is inverted (top row). This is because overmassive BHs are correlated with higher BH luminosities (see Figure \ref{fig:PhysicalPropsBHStellar}), and this correlation is significantly stronger than the one for the outflow rates. 

Overall, we find that the trends at fixed gas mass from Section \ref{subsec:offsetBHs} are still recovered at fixed stellar mass. An exception to this are the mass loading factor and the sSFR. At low redshifts, we only see a clear trend for the upper end of the massive dwarf regime ($M_\mathrm{stellar} \gtrsim 6 \times 10^{9} \ \Msun$) and the $\mathcal{M^{*}}$ galaxies. This difference is most likely due to the strong SN feedback in \fable \ which clears the gas out of dwarf galaxies, preventing the AGN from accreting efficiently. However, at higher redshifts ($z \geq 2$), where accretion rates are higher and a significant fraction of BHs is in the quasar mode (see Figure~\ref{fig:PopulationStatistics}), there is a clear correlation between overmassive BHs and suppressed sSFRs across the whole stellar mass range (see Figure~\ref{fig:sSFR_overmassive_BHs}). This indicates that if the AGN are able to accrete efficiently, they are able to regulate star formation in dwarf galaxies.


\bsp	
\label{lastpage}
\end{document}